\documentclass[10pt,a4paper,twoside]{article}
\usepackage[english]{babel}
\usepackage[utf8]{inputenc}
\usepackage{amssymb}
\usepackage{amsmath}
\usepackage{fancyhdr}
\usepackage{verbatim}
\usepackage{lastpage}
\usepackage{ifthen}

\usepackage{booktabs}
\usepackage{tabularx}
\usepackage{multirow}
\usepackage{adjustbox}
\usepackage{subcaption}
\usepackage[numbers]{natbib}
\usepackage{float}

\usepackage[pdftex]{color}
\usepackage{graphicx}
\usepackage[hidelinks]{hyperref}
\usepackage{iberamia}   

\newcommand{\thispaperdoi}{}
\title{Credit Risk Meets Large Language Models: Building a Risk Indicator from Loan Descriptions in P2P Lending}

\author{Mario Sanz-Guerrero$^{[1,2,*]}$, Javier Arroyo$^{[3,4]}$
\smallskip \\
\small{
$^{[1]}$Facultad de Informática, Universidad Complutense de Madrid, Spain\\
$^{[2]}$Johannes Gutenberg University Mainz, Germany\\
$^{[3]}$Instituto de Tecnología del Conocimiento, Universidad Complutense de Madrid, Spain\\
$^{[4]}$Departamento de Ciencias de la Computación, Universidad de Alcalá de Henares, Spain\\
\smallskip
$^{[*]}$msanzgue@uni-mainz.de\\
}}
\date{}

\begin{document}

\medskip
\maketitle

\pagestyle{Rest}
\thispagestyle{FirstPage}
\smallskip

\noindent{\small{\textbf{Abstract}
Peer-to-peer (P2P) lending connects borrowers and lenders through online platforms but suffers from significant information asymmetry, as lenders often lack sufficient data to assess borrowers' creditworthiness. This paper addresses this challenge by leveraging BERT, a Large Language Model (LLM) known for its ability to capture contextual nuances in text, to generate a risk score based on borrowers' loan descriptions using a dataset from the Lending Club platform. We fine-tune BERT to distinguish between defaulted and non-defaulted loans using the loan descriptions provided by the borrowers. The resulting BERT-generated risk score is then integrated as an additional feature into an XGBoost classifier used at the loan granting stage, where decision-makers have limited information available to guide their decisions. This integration enhances predictive performance, with improvements in balanced accuracy and AUC, highlighting the value of textual features in complementing traditional inputs. Moreover, we find that the incorporation of the BERT score alters how classification models utilize traditional input variables, with these changes varying by loan purpose. These findings suggest that BERT discerns meaningful patterns in loan descriptions, encompassing borrower-specific features, specific purposes, and linguistic characteristics. However, the inherent opacity of LLMs and their potential biases underscore the need for transparent frameworks to ensure regulatory compliance and foster trust. Overall, this study demonstrates how LLM-derived insights interact with traditional features in credit risk modeling, opening new avenues to enhance the explainability and fairness of these models.}}
\medskip



\section{Introduction}
\label{sec:introduction}
Peer-to-peer (P2P) lending is a growing phenomenon that allows individuals to engage in direct lending and borrowing transactions, bypassing traditional financial institutions. The process is facilitated through online platforms, where prospective borrowers submit loan applications and potential lenders make informed decisions about where to invest their funds.

An inherent challenge in P2P lending is the presence of information asymmetry, wherein borrowers possess more and often superior information compared to lenders. To address this issue, platforms employ strategies to complement the conventional data provided in loan applications \cite{Cummins2019}. For instance, borrowers are frequently encouraged to provide a voluntary textual description describing the purpose of the loan and their particular situation. Despite the absence of formal verification, such voluntary disclosures have been observed to stimulate increased bidding activity among lenders. However, lenders may lack the expertise to assess the creditworthiness of borrowers effectively and may be influenced by different factors \cite{Michels2012}.

Traditional credit scoring models often overlook the valuable information contained in loan applicants' narratives \cite{MONTEVECHI2024}. Several methods have attempted to incorporate this data, including extracting linguistic metrics \cite{GaoLin2015_lemon, DORFLEITNER2016}, using topic modeling to identify underlying themes \cite{Xia2020, Zhang2020}, or combining both approaches \cite{Siering2023}. These methods offer clear advantages, such as computational efficiency---particularly in the case of linguistic metrics---and greater interpretability. However, they also come with drawbacks, including limited ability to capture meaning---especially for linguistic metrics---, dependence on extensive preprocessing in the case of topic modeling, and an overall inability to fully grasp nuance and context.

In contrast, transformer-based models, such as BERT (Bidirectional Encoder Representations from Transformers) \cite{BERT2018}, offer a powerful alternative. While these models are less interpretable, they excel at understanding text by capturing semantic and contextual relationships at the word, sentence, and document levels. They also adapt flexibly to variations in style and structure. Moreover, leveraging pre-trained models is straightforward---by transferring general language knowledge from large corpora, these models can be fine-tuned for specific tasks or domains, achieving significant performance improvements with relatively little labeled data.

In sum, BERT's bidirectional training and context-sensitive representations make it well-suited for tasks requiring deep semantic understanding. It has been successfully fine-tuned for various classification tasks \cite{Sun_2019_BERTfinetune}, including applications in biomedicine \cite{BioBERT} and specialized areas such as spam detection \cite{tida2022}. Recently, Xia et al.\ \cite{Xia2023} demonstrated the effectiveness of a fine-tuned BERT model in discriminating P2P loans within the Chinese market.

In this study, we extend this line of research by applying BERT at the loan granting stage---a critical point where customer narratives hold greater importance due to the limited information available for decision-making. As our baseline method, we employ XGBoost, an efficient gradient-boosting algorithm that builds ensembles of decision trees to enhance performance and that has demonstrated its ability in loan granting scenarios \cite{ArizaExplainability}. We show that incorporating BERT-generated risk scores into a loan granting model significantly enhances predictive performance, and we delve deep into how BERT processes textual data and influences model behavior. Specifically, our analysis reveals that BERT captures a wide range of information from loan descriptions, including borrower-specific attributes, loan purposes, and linguistic features embedded in the text. Furthermore, we demonstrate that the inclusion of BERT-generated scores reshapes how credit models leverage other input variables, with the impact varying substantially across different loan purposes. While our findings highlight BERT's potential to improve credit risk assessment in P2P lending, they also emphasize the importance of transparency in understanding what these models learn from text. Such transparency is crucial for building trust among stakeholders and ensuring acceptance by entities.

This paper is organized as follows: Section \ref{sec:related} provides a comprehensive review of related work in credit risk assessment and natural language processing. Section \ref{sec:BERT} presents an overview of LLMs and the BERT model. Section \ref{sec:Dataset_des} describes the dataset used, detailing the data preprocessing steps and conducting an in-depth exploratory data analysis. Section \ref{sec: DescExperiment} outlines the methodology, model architecture, and training procedures employed in integrating BERT into the credit risk assessment framework. Section \ref{sec:BERTscoreAnalysis} analyzes the risk score generated by the BERT description processing. Section \ref{sec: Classification Results} discusses the results of our experiments, highlighting the improvements achieved by incorporating BERT-based textual analysis. Finally, Section \ref{sec: Conclusion} concludes the paper by summarizing key findings, discussing implications, and suggesting avenues for future research in the intersection of NLP and credit risk assessment.

\section{Related Work}
\label{sec:related}

\subsection{Data Sources in Credit Risk Modeling: The Use of Loan Descriptions}
In their comprehensive analysis of risk-return modeling within the P2P lending market \cite{Miller2021}, the authors identify a discernible trend towards including new sources and types of information to improve risk and profit management models in the P2P market. The sources are very diverse and include transactional data \cite{Xu2022}, the topology of the lending-borrowing network \cite{LI2018}, data from social networks \cite{Xu2016}, or, more recently, facial features \cite{Qi2022}. Among them, the authors identify as a predominant trend the inclusion of textual data taken from statements describing the purpose of the loan.

In a pioneer work exploring the impact of textual factors on peer-to-peer lending \cite{Herzenstein2011}, the authors analyze P2P loans including manually annotated narrative aspects, such as trustworthiness, economic hardship, hard work, success, morality, and religiosity. These aspects were combined with demographic variables and loan characteristics. Their results highlight that narratives regarding trustworthiness strongly influence decision-makers, particularly credit lenders, in their loan approval process. Additionally, some of these narratives play a substantial role in subsequent loan performance.

However, most subsequent studies typically use text mining or artificial intelligence methods to extract linguistic features or loan description topics. Regarding the use of linguistic features, the authors in \cite{GaoLin2015_lemon} use machine learning and text mining techniques to quantify and extract linguistic features (e.g., readability, positivity, objectivity, and deception cues), and then build both explanatory econometric models and predictive models using such features. They find that they can indeed reflect borrowers' creditworthiness and predict loan default. They also use a panel of investors and confirm that investors indeed value texts written by borrowers, but that they can also be deceived by some of the deception cues well established in the literature. Similarly, in \cite{DORFLEITNER2016}, the authors include linguistic factors and the presence of social and emotional keywords and evaluate their impact on two European platforms. They found that text-derived variables influence the probability of funding, but not the probability of default. In \cite{Wang2016}, linguistic statistical features and abstract text features (including deception, subjectivity, sentiment, readability, personality, and mindset) are used to characterize text descriptions. They compare the performance of different classifiers based on the textual features and conclude that their performance is close to that of the classifiers using traditional financial features, but that adding textual features can improve the performance of the whole credit risk evaluation system.

\subsection{Topic Modeling Approaches}
As for topic modeling, the Latent Dirichlet Allocation model (LDA) has been widely used. In \cite{Jiang2017}, the authors use LDA to extract six topics from the loan descriptions whose meanings are obvious: assets, income and expenses, work, family, business, and agriculture. They also consider the number of characters in the descriptive text. They conclude that soft (qualitative) information can improve the performance of loan default prediction compared to existing methods based only on hard (quantitative) information and that soft features have a significant ability to discriminate loan defaults. Similarly, in \cite{Yao2018}, an LDA topic model is used to classify the loan titles into six purposes. Their findings reveal that the stated purpose significantly influences a borrower's chances of securing financing. Notably, ambiguous titles---where borrowers fail to clearly articulate the loan's purpose---substantially diminish the likelihood of loan approval. In \cite{Xia2020}, Xia et al.\ used a keyword clustering algorithm for automatic topic extraction. Their method combines keyword extraction based on term frequency-inverse document frequency (TF-IDF) with word embeddings generated by the Word2Vec neural network model \cite{word2vec}. Analysis of three real-world datasets demonstrated that incorporating these topic variables significantly enhanced predictive accuracy compared to relying solely on traditional information.

Siering \cite{Siering2023} recently examined the impact of both topical and linguistic features on loan default prediction. To extract topics, the author employed a financial text analysis method \cite{LOUGHRAN_MCDONALD2011} to construct a domain-specific dictionary. The identified topics captured elements such as the loan purpose, the borrower’s requests for assistance, expressions of reliability, and appreciation. These topics were represented as binary indicator variables. Additionally, text mining techniques were used to generate features measuring attributes like polarity, \textit{active orientation}, readability, average sentence length, and word count. These features were then incorporated into a logistic regression model, revealing that both linguistic and content-based factors contribute to predicting loan default probability, with content-based factors showing greater significance. The analysis further indicated that certain variables, such as expressions of reliability, positively correlate with loan repayment likelihood.

\subsection{Advancements with Large Language Models}
In recent years, several studies have begun to explore state-of-the-art natural language processing techniques, including deep learning models, for loan default prediction. In their work \cite{Zhang2020}, Zhang et al.\ investigated transformer encoders for extracting textual features from loan descriptions. These features, combined with traditional hard features from loan applications, were input into a neural network to predict default probabilities. Their findings demonstrated the effectiveness of transformers, with models incorporating textual loan descriptions outperforming those that did not.

More recently, Xia et al.\ \cite{Xia2023} explored the use of BERT-based LLMs to extract information from loan narratives, integrating these insights into both logistic regression and machine learning models like random forest and deep forest. While this work achieved improved predictive performance, it lacked transparency regarding the specific information captured by the LLM and how it influenced the classification model, leaving the results largely opaque and black-box.

\subsection{Research Gap}
Despite the progress made in leveraging textual data for credit risk modeling, several gaps remain. First, while advanced NLP techniques like BERT have been applied to loan descriptions, the interpretability of the extracted features and their influence on downstream models remains underexplored. For instance, a previous work \cite{Xia2023} achieved improved performance but provided limited insights into how the BERT-derived features contributed to the model's predictions. Second, there is a need for a more systematic integration of textual features with traditional financial variables in a way that enhances both predictive accuracy and interpretability.

Our study addresses these gaps by applying BERT to create a risk score from loan descriptions in the loan-granting process, a context where customer narratives play a crucial role. Unlike previous work, we analyze the score, including its relationship with other variables, its impact on the classification model, and on the results obtained across the predefined loan purposes segmented, aiming to provide more interpretability and insights into the decision-making process.

\section{Basics of LLM architectures and BERT}
\label{sec:BERT}

Large Language Models (LLMs) are built upon the Transformer architecture \cite{Transformer2017}, leveraging attention mechanisms to enhance language comprehension. LLMs can be broadly categorized into three primary families, each distinguished by its architecture: 
\begin{itemize}
    \item Encoder-only, widely employed for language comprehension tasks such as text classification, named entity recognition, and extractive question answering. The most famous example is BERT \cite{BERT2018}, which will be explained in more detail below.
    \item Decoder-only, designed for generative tasks, exemplified by the well-known GPT models \cite{gpt2, gpt3}. It is employed in various tasks, including question answering \cite{pride2023core}, text summarization \cite{bhaskar2022prompted}, and programming code generation \cite{chen2021evaluating}.
    \item Encoder-decoder models, suited for tasks demanding both language understanding and generation, such as language translation or text summarization. The most influential models are BART \cite{BART2019} and T5 \cite{T52020}. 
\end{itemize}

The selection of the appropriate architecture hinges on the specific requirements of the intended task. Whether it be the nuanced comprehension of language, creative text generation, or the synthesis of both, the versatility of LLMs offers a tailored solution for diverse applications.

We will focus on BERT (Bidirectional Encoder Representations from Transformer), which is a Trans\-former-\-based language model introduced by Google researchers in 2018 \cite{BERT2018}. BERT's architecture consists of a stack of encoders from the Transformer model. The bidirectional nature of BERT is key, as it considers both the left and right context of each word, enhancing its ability to understand context-dependent meanings and to be effective in language understanding tasks. Numerous studies have consistently shown that BERT is the most effective linguistic model for various of these tasks \citep{Kriebel2022, Stevenson2021}. Notably, BERT has 340 million parameters, while the widely recognized GPT-3 model has 175 billion, making BERT 514 times smaller than GPT-3 \cite{gpt3}. Given this significant size difference, BERT can be operated on standard home equipment for model inference, which greatly simplifies its use in practical scenarios. In contrast, GPT, built with a Transformer decoder stack, not only demands much more powerful equipment but is suited for language generation tasks.

BERT stands as a milestone whose success has spurred the development of a diverse family of models that build upon its architecture. Some versions aim to achieve similar performance while having a smaller number of parameters, such as DistilBERT (a distilled version of BERT) \cite{DistilBERT}, or ALBERT (A Little BERT) \cite{ALBERT}. Others are adaptations to other languages such as CamemBERT \cite{martin-etal-2020-camembert} to French or BETO to Spanish \cite{BETO}. Other proposals aimed to improve upon BERT by modifying some design decisions when pretraining BERT and also training the model longer, as in the case of RoBERTa (Robustly optimized BERT approach) \cite{RoBERTa}, which resulted in improved contextualized representations and enhanced language understanding.

To further elucidate the role of BERT in specialized applications, it is crucial to understand its capacity for transfer learning and fine-tuning. Transfer learning involves using a pre-trained model like BERT, which has initially learned general language patterns from a large corpus to a specific task or dataset. This technique allows us to take advantage of the rich linguistic representations without needing extensive computation from scratch. Fine-tuning involves adjusting the pre-trained model's parameters to capture the nuances of the target task or application field by further training with new instances from the new context. For example, BioBERT is a BERT model fine-tuned for biomedical text mining tasks like named entity recognition and question answering \cite{BioBERT}. Other adaptations have targeted text classification and sentiment analysis in specific datasets \cite{Sun_2019_BERTfinetune, GAO2019}. Finally, as already mentioned, \cite{Xia2023} shows how fine-tuned Chinese BERT models can enhance the classification performance of default loans in the P2P market.

\section{Dataset}
\label{sec:Dataset_des}

We use a public data set of the P2P lending company Lending Club\footnote{\url{https://www.kaggle.com/wordsforthewise/lending-club}}, which is widely used in credit risk publications and the most widely used when dealing with the P2P market \cite{ABASHA2021, Miller2021}. However, instead of using the original dataset, which includes 2,260,699 loans granted by the company between 2007 and 2018, we use a version modified for proposing granting models \cite{DatasetAriza}, used in \cite{ArizaExplainability, Miller2024}. Since granting models determine which loans will be fully repaid, its estimation needs loans whose final status is known (i.e., that were either fully repaid or defaulted). Thus, the dataset excludes loans in transitory states (in a grace period, late, etc.) and loans with no information on income and indebtedness, which is essential to compute the input variables, resulting in 1,347,681 instances.

Additionally, the original dataset contains variables detailing the loan's lifecycle and other post-application aspects (e.g., the interest rate). In contrast, our version only includes variables available at the time of application, which are those utilized by granting models.

Loan descriptions were inconsistently available, appearing only for certain loans between April 2008 and March 2014. To accurately assess the impact of textual descriptions on default prediction, our analysis focuses solely on the 119,101 loans that include the \textit{desc} variable. Kolmogorov-Smirnov and chi-square tests were applied to quantitative and categorical variables to assess potential bias from filtering. The lack of significant differences indicates that the filtered dataset is representative of the original dataset.

\begin{table}[]
    \small
    \caption{Variable description.}\label{tab:var_desc}
    \begin{tabularx}{\textwidth}{@{}lX@{}}
        \toprule
            Variable & Description \\
        \midrule
        \textbf{Quantitative variables} & \\
            \textit{revenue} & Borrower's self-declared annual income during registration. \\
            \textit{dti\_n} & Indebtedness ratio for obligations excluding mortgage. Monthly information. \\
            \textit{loan\_amnt} & Amount of credit requested by the borrower. \\
            \textit{fico\_n} & Credit bureau score. Defined between 300 and 850, reported by Fair Isaac Corporation as a summary risk measure based on historical credit information reported at the time of application. \\
        \midrule
        
        \textbf{Categorical variables} & \\
            \textit{emp\_length} & Employment length of the borrower categorized into 12 categories, including the no information category. \\
            \textit{purpose} & Credit purpose category for the loan request. \\
            \textit{home\_ownership} & Homeownership status provided by the borrower. \\
            \textit{addr\_state} & Borrower's residence state from the USA. \\
        \midrule
        
        \textbf{Textual variable} & \\
            \textit{desc} & Description of the credit request provided by the borrower. \\
        \bottomrule
    \end{tabularx}
\end{table}

In the dataset, the target variable suffers the usual class imbalance problem (only 15.27\% of default), which will be considered in the design of the experiments. Table \ref{tab:var_desc} shows the input variables of our granting model, which are explained below.

As for the quantitative variables, the Fair Isaac Corporation credit bureau (FICO) information in the original dataset is given by a minimum and maximum range of limits to which the borrower's FICO belongs at loan origination. However, we average these two values to have a single indicator of the creditworthiness of potential borrowers resulting in our \textit{fico\_n} variable. For the case of the debt variable, \textit{dti\_n} is estimated from the original dataset variables as the ratio calculated from the total debts of the co-borrowers over the total debt obligation divided by the combined monthly income of the co-borrowers.

Regarding the categorical variables, we merged the categories `other', `none', and `any' into a unified category labeled `other' for the \textit{home\_ownership} variable. This decision was made due to a lack of clear differentiation among these options, coupled with their similar default percentages and their relatively low percentages of occurrences. The \textit{emp\_length} variable was treated as categorical rather than numerical since it includes categories for `no information' and for `more than ten years'.

For the textual variable, we carried out an exhaustive work of text cleaning. First, we removed all those descriptions that contained the default description provided by Lending Club on its web form (\textit{``Tell your story. What is your loan for?''}). Moreover, we removed the prefix \textit{``Borrower added on DD/MM/YYYY $>$''} from the descriptions, as we did not want any temporal background on them. Finally, as these descriptions came from a web form, we replaced all HTML entities with their corresponding characters (e.g. `\&amp;' was substituted by `\&', `\&lt;' was substituted by `$<$', etc.).

\begin{table}[]
    \centering
    \small
    \caption{Exploratory data analysis. Quantitative variables.}\label{tab:eda_quant}
    \begin{tabular}{llrrr}
        \toprule%
            Variable & Statistic & Non-Default & Default & Total \\
        \midrule
            \multirow{5}{*}{revenue} & Mean & \$ 73,570.69 & \$ 66,218.66 & \$ 72,447.84 \\\cmidrule{2-5}
             & Median & \$ 64,000.00 & \$ 58,000.00 & \$ 62,000.00 \\\cmidrule{2-5}
             & SD & \$ 54,944.60 & \$ 40,731.98 & \$ 53,086.79 \\\cmidrule{2-5}
             & KS D-Test & & & 0.09* \\
        \midrule
            \multirow{5}{*}{dti\_n} & Mean & 16.15 & 17.67 & 16.38 \\\cmidrule{2-5}
             & Median & 15.88 & 17.70 & 16.16\\\cmidrule{2-5}
             & SD & 7.50 & 7.53 & 7.53 \\\cmidrule{2-5}
             & KS D-Test & & & 0.09* \\
        \midrule
            \multirow{5}{*}{loan\_amnt} & Mean & \$ 13,799.25 & \$ 15,111.44 & \$ 13,999.66 \\\cmidrule{2-5}
             & Median & \$ 12,000.00 & \$ 14,000.00 & \$ 12,000.00 \\\cmidrule{2-5}
             & SD & \$ 7,931.55 & \$ 8,363.19 & \$ 8,012.86 \\\cmidrule{2-5}
             & KS D-Test & & & 0.08* \\
        \midrule
            \multirow{5}{*}{fico\_n} & Mean & 705.22 & 694.20 & 703.54 \\\cmidrule{2-5}
             & Median & 697.00 & 687.00 & 697.00 \\\cmidrule{2-5}
             & SD & 33.32 & 27.48 & 32.74 \\\cmidrule{2-5}
             & KS D-Test & & & 0.15* \\
        \bottomrule
        \multicolumn{5}{l}{\footnotesize * p-value less than 0.01.}
    \end{tabular}
\end{table}

Table \ref{tab:eda_quant} presents the quantitative variables and the results of the Kolmogorov-Smirnov test, which was used to compare the empirical cumulative distribution functions of Default and Non-default loans. According to these results, defaulted loans are characterized by lower revenue, higher debt-to-income ratio (\textit{dti\_n}), higher requested amount (\textit{loan\_amnt}), and lower FICO scores (\textit{fico\_n}), being the differences significant at the 0.01 level.

\begin{table}[]
    \centering
    \small
    \caption{Exploratory data analysis. Categorical variables.}\label{tab:eda_cat}
    \begin{tabular}{llrrrr}
        \toprule%
            Variable & Category & Count & Rel. Freq. & Default Rate & Chi Test \\
        \midrule
            \multirow{5}[2]{*}{home\_ownership} & MORTGAGE & 60,796 & 51.05\% & 14.14\% & \multirow{5}[2]{*}{131.08*} \\\cmidrule{2-5}
             & OTHER & 143 & 0.12\% & 20.98\% & \\\cmidrule{2-5}
             & OWN & 9,582 & 8.05\% & 15.69\% & \\\cmidrule{2-5}
             & RENT & 48,580 & 40.79\% & 16.60\% & \\
        \midrule
            \multirow{15}[10]{*}{emp\_lenght} & $<$ 1 year & 9,548 & 8.02\% & 14.83\% & \multirow{15}[10]{*}{104.96*} \\\cmidrule{2-5}
             & 1 year & 7,803 & 6.55\% & 14.39\% & \\\cmidrule{2-5}
             & 2 years & 10,960 & 9.20\% & 14.68\% & \\\cmidrule{2-5}
             & 3 years & 9,370 & 7.87\% & 14.18\% & \\\cmidrule{2-5}
             & 4 years & 7,561 & 6.35\% & 14.56\% & \\\cmidrule{2-5}
             & 5 years & 9,019 & 7.57\% & 14.76\% & \\\cmidrule{2-5}
             & 6 years & 7,271 & 6.10\% & 16.04\% & \\\cmidrule{2-5}
             & 7 years & 6,638 & 5.57\% & 15.59\% & \\\cmidrule{2-5}
             & 8 years & 5,374 & 4.51\% & 15.39\% & \\\cmidrule{2-5}
             & 9 years & 4,356 & 3.66\% & 15.79\% & \\\cmidrule{2-5}
             & 10+ years & 36,287 & 30.47\% & 15.41\% & \\\cmidrule{2-5}
             & NI & 4,914 & 4.13\% & 19.78\% & \\
        \midrule
            \multirow{19}[9]{*}{purpose} & car & 1,884 & 1.58\% & 9.61\% & \multirow{19}[9]{*}{568.47*} \\\cmidrule{2-5}
             & credit card & 25,051 & 21.03\% & 12.81\% & \\\cmidrule{2-5}
             & debt consolidation & 68,372 & 57.41\% & 16.14\% & \\\cmidrule{2-5}
             & educational & 265 & 0.22\% & 16.98\% & \\\cmidrule{2-5}
             & home improvement & 7,170 & 6.02\% & 12.93\% & \\\cmidrule{2-5}
             & house & 805 & 0.68\% & 15.78\% & \\\cmidrule{2-5}
             & major purchase & 3,062 & 2.57\% & 10.65\% & \\\cmidrule{2-5}
             & medical & 970 & 0.81\% & 17.32\% & \\\cmidrule{2-5}
             & moving & 768 & 0.64\% & 14.84\% & \\\cmidrule{2-5}
             & other & 6,361 & 5.34\% & 17.69\% & \\\cmidrule{2-5}
             & renewable energy & 127 & 0.11\% & 19.69\% & \\\cmidrule{2-5}
             & small business & 2,518 & 2.11\% & 26.41\% & \\\cmidrule{2-5}
             & vacation & 561 & 0.47\% & 16.22\% & \\\cmidrule{2-5}
             & wedding & 1,187 & 1.00\% & 12.47\% & \\
        \bottomrule
        \multicolumn{6}{l}{* p-value less than 0.01.}
    \end{tabular}
\end{table}

Similarly, Table \ref{tab:eda_cat} displays the distribution of categories within each categorical variable and the corresponding default rates. The \textit{addr\_state} variable is excluded due to its 50 categories, one for each U.S. state. The table also indicates whether there is a significant dependence between the target variable and the categorical variables at the 0.01 significance level. The results show significant dependence for all variables, including the \textit{addr\_state} variable not reported in the table (test value of 211.12).

In the \textit{home\_ownership} variable, the `OTHER' category shows the highest risk (20.98\%) but a small frequency (0.12\%), while the `MORTGAGE' category is the most frequent (51.05\%) and the least risky (14.14\%) one. In the \textit{emp\_length} variable, the category that denotes no information (`NI') has the highest risk (19.78\%), but also the lowest frequency (4.13\%). In general, employment length can be categorized into two groups with comparable default rates: those with employment lengths of five years or less and those with more than five years. Interestingly, the risk is slightly higher in the group with more than five years of employment. The categories are not perfectly ordered, which supports the use of one-hot encoding to treat this variable as categorical. Finally, the most frequent \textit{purpose} is `debt consolidation', constituting 57\% of the loans, which has a default rate of 16.14\%. Notably, the riskiest purpose is `small business', with a 26.41\% default rate. Conversely, `car' loans demonstrate the lowest risk, with a mere 9.61\% default rate. This striking divergence in default rates across diverse purposes underscores a significant variability in risk within the various loan purposes.

\begin{table}[]
    \centering
    \small
    \caption{Exploratory data analysis. Textual variable (\textit{desc}).}\label{tab:eda_text}
    \begin{tabular}{llrrr}
        \toprule%
            Variable & Statistic & Non-Default & Default & Total \\
        \midrule
            \multirow{5}{*}{Word count} & Mean & 36.72 & 35.49 & 36.54 \\\cmidrule{2-5}
             & Median & 24.0 & 22.0 & 24.0 \\\cmidrule{2-5}
             & SD & 46.62 & 48.78 & 46.96 \\\cmidrule{2-5}
             & KS D-Test &  &  & 0.03* \\
        \midrule
            \multirow{5}{*}{Readability} & Mean & 66.70 & 66.31 & 66.64 \\\cmidrule{2-5}
             & Median & 73.88 & 74.19 & 74.02 \\\cmidrule{2-5}
             & SD & 32.87 & 35.55 & 33.29 \\\cmidrule{2-5}
             & KS D-Test &  &  & 0.02* \\
        \midrule
            \multirow{5}{*}{Polarity} & Mean & 0.0964 & 0.0909 & 0.0956 \\\cmidrule{2-5}
             & Median & 0.0367 & 0.0 & 0.0320 \\\cmidrule{2-5}
             & SD & 0.1685 & 0.1699 & 0.1687 \\\cmidrule{2-5}
             & KS D-Test &  &  & 0.04* \\
        \midrule
            \multirow{5}{*}{Subjectivity} & Mean & 0.3193 & 0.3029 & 0.3168 \\\cmidrule{2-5}
             & Median & 0.3635 & 0.3333 & 0.3589 \\\cmidrule{2-5}
             & SD & 0.2542 & 0.2595 & 0.2551 \\\cmidrule{2-5}
             & KS D-Test &  &  & 0.04* \\
        \bottomrule
        \multicolumn{5}{l}{\footnotesize * p-value less than 0.01.}
    \end{tabular}
\end{table}

Regarding the textual description of the loan (\textit{desc} variable), Table \ref{tab:eda_text} shows some metrics to characterize it. There is a one-word difference in the average word count between the descriptions of defaulted and non-defaulted obligations. The readability was calculated using the Flesch Reading Ease Score\footnote{Calculated with Textstat (Python library). Source: \url{https://github.com/textstat/textstat}.}, which indicates the approximate educational level required for comfortable comprehension of a given text (higher scores denote greater ease of reading). The texts in both categories have scores around 66, signifying that they can be readily comprehended by students aged 13 to 15. Additionally, we analyzed the average polarity and average subjectivity\footnote{Calculated with TextBlob (Python library). Source: \url{https://github.com/sloria/textblob}.}. The polarity, ranging from -1 to 1 to denote negative or positive sentiment, was observed to be approximately 0.1 in both cases, suggesting a subtle positive sentiment. On the other hand, subjectivity, measuring the presence of judgments and opinions on a scale from 0 to 1, exhibited values close to 0.31 in both categories. This indicates that while the texts in both cases maintain a generally objective tone, there is a discernible inclusion of some judgments or opinions.

Although the distinctions in these metrics between the default and non-default categories are subtle, their significance is confirmed by the Kolmogorov-Smirnov test. Consequently, it is pertinent to incorporate linguistic aspects into credit risk modeling. Our approach for extracting information from the descriptions relies on leveraging LLMs capable of encompassing not just linguistic nuances but also capturing content details. We elaborate on our methodology below.

\section{Methodology}
\label{sec: DescExperiment}

This study employs transfer learning to enable an LLM to generate a score that reflects the likelihood of loan default based on textual descriptions. We explore how a fine-tuned LLM captures various aspects of loan descriptions that are indicative of default risk. Furthermore, we demonstrate that incorporating the LLM-generated score enhances the predictive accuracy of a loan-granting model and fundamentally alters how the model operates compared to when the score is absent.

Our baseline model is a machine learning classifier that utilizes all available variables from the loan application process, including both quantitative and categorical data. Specifically, we use XGBoost, which has been shown to deliver superior performance in similar loan-granting contexts \cite{ArizaExplainability}. To augment this baseline, we introduce the LLM-generated default risk score as an additional input feature. We analyze the information captured by this score and assess its impact on both prediction performance and the behavior of the resulting model.

\begin{figure}[]
    \centering
    \includegraphics[width = 0.8\textwidth]{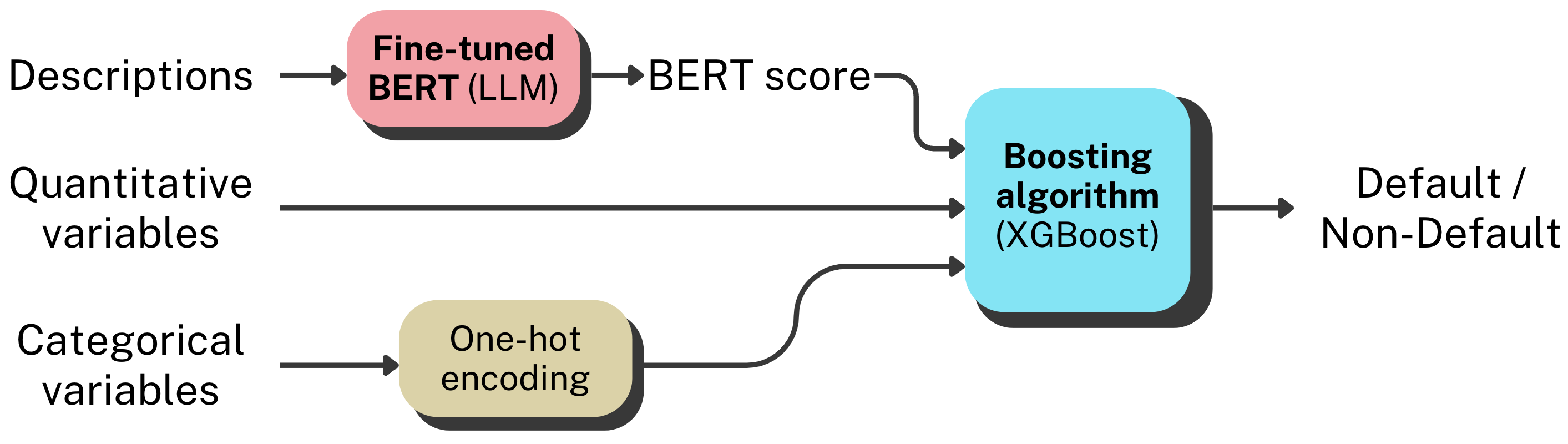}
    \caption{Diagram of the experiment architecture.}
    \label{fig:experiment_architecture}
\end{figure}

The experimental setup is shown in Figure \ref{fig:experiment_architecture}. Loan descriptions are processed by a fine-tuned BERT model, which outputs a \textit{BERT\_score} representing the probability of default. This score is then integrated with other input variables in the XGBoost classifier to produce the final prediction.

The key components of the methodology are elaborated upon in this section.

\subsection{Tuning the classifier}
\label{ssec:Tuning the classifier}

The classification algorithm used was XGBoost \cite{XGBoost}, which is trained using a stratified $k$-fold cross-validation, dividing the dataset into $k$ subsets (here, $k = 5$) and preserving the original class distribution in each fold to provide a more reliable evaluation in an imbalanced dataset as ours. 
The dataset is shuffled to eliminate biases derived from its original ordering and to ensure representative subsets, avoiding artificial patterns in the training. Furthermore,  k=5 is used to balance the reduction of variance in the estimates with the need for sufficiently large partitions, a crucial factor in datasets with few instances and high heterogeneity as ours.

Furthermore, the instances are shuffled to avoid potential ordering biases in the dataset. 
To fine-tune the hyperparameters we used a genetic algorithm \cite{holland1992adaptation} to evolve candidate hyperparameter combinations and choose the one that maximizes the fitness measure, which was the average balanced accuracy (BACC) in the 5 validation sets of the cross-validation. We use BACC as it accounts for the imbalanced nature of the dataset. In preliminary experiments, we also considered the area under the receiver operating characteristic (AUROC), which measures the model's ability to discriminate between positive and negative examples regardless of the classification threshold chosen. However, we observed that the resulting XGBoost classifiers produced poor BACC values (similar to those from a na\"ive classifier that predicts the majority class) when using the standard 0.5 threshold to make the prediction. We also observed that XGBoost classifiers with extremely similar AUROC values produced very different results in terms of BACC. Thus, we decided to use the BACC measure as it resulted in classifiers with more stable behavior.

In genetic optimization, each individual is characterized by its genes, that is, the considered hyperparameters of the XGBoost. 
We have included several kinds of parameters, including:
\begin{itemize}
    \item Parameters that adjust the sample weights, such as \textit{scale\_pos\_weight}, which balance the weights of the classes and are useful in unbalanced datasets as ours.
    \item Parameters that set up the behavior of the boosting algorithm, such as the learning rate (\textit{eta}), the percentage of subsamples in each iteration (\textit{subsample}), the number of learners (\textit{n\_estimators}), and the percentage of dataset features that use each learner (\textit{colsample\_bytree}).
    \item Parameters that control the learning process of each tree, including its maximum depth (\textit{max\_depth}), regularization parameters (\textit{lambda} and \textit{alpha}), the loss reduction required to make a further partition on a leaf node (\textit{gamma}), and the minimum number of weighted instances needed in a child node (\textit{min\_child\_weight}).
\end{itemize}

Table \ref{chromosomes} shows the hyperparameters together with their respective ranges, which were determined based on a combination of domain knowledge, preliminary experiments, and established practices in the literature \cite{ArizaExplainability,Ye2018genetic}. 
For instance, the learning rate (\textit{eta}) was set between 0.001 and 0.5 to balance the trade-off between convergence speed and model performance. The \textit{max\_depth} parameter, which controls the maximum depth of a tree, was set between 2 and 12 to prevent overfitting while allowing sufficient model complexity. Similarly, the ranges for other hyperparameters, such as regularization terms and sampling rates, were selected to ensure a wide exploration of their effects, which are crucial for optimizing model performance in imbalanced datasets.

\begin{table}[]
    \centering
    \small
    \caption{Hyperparameters of XGBoost considered in the genetic optimization and their respective ranges.}\label{chromosomes}%
    \begin{tabular}{@{}lrr@{}}
        \toprule
            Parameter & Min. value & Max. value \\
        \midrule
            \textit{scale\_pos\_weight} & 0.1 & 10 \\
            \textit{eta} (learning rate) & 0.001 & 0.5 \\
            \textit{subsample} & 0.7 & 1 \\
            \textit{n\_estimators} & 2 & 500 \\
            \textit{colsample\_bytree} & 0.3 & 1 \\
            \textit{max\_depth} & 2 & 12 \\
            \textit{lambda} & 0.5 & 10 \\
            \textit{alpha} & 0.5 & 10 \\
            \textit{gamma} & 0 & 10 \\            
            \textit{min\_child\_weight} & 0 & 10 \\
        \bottomrule
    \end{tabular}
\end{table}

The evolutionary strategy chosen for the optimization is the ``\textit{Mu} plus \textit{lambda}'' ($\mu + \lambda$) approach, where $\mu$ represents the number of individuals to select for the next generation, and $\lambda$ indicates the number of children to produce at each generation. Unlike traditional approaches where children often replace parents, the $\mu + \lambda$ strategy involves adding both children and parents to produce the next generation. This strategy was selected to maintain diversity in the population and prevent premature convergence to suboptimal solutions. In the context of this research, we set $\mu$ to 150 and $\lambda$ to 150. This configuration, chosen based on empirical testing, provides a balance between exploration of the search space and computational feasibility. Increasing $\mu$ and $\lambda$ beyond these values resulted in marginal improvements at significantly higher computational costs.

Initially, pairs of parents are chosen through tournament selection with a tournament size of 2. Subsequently, the children are generated employing a two-point crossover technique on the parents' chromosomes with an 80\% probability and applying a random resetting mutation with a 20\% probability. The random resetting mutation implies that each gene of every child has a 20\% chance of acquiring a new random value within its defined range. These probabilities were chosen based on preliminary experiments that demonstrated a good balance between exploration and exploitation. To create an offspring of 150 children, this selection, crossover, and mutation process is repeated 75 times. Finally, we combine the 150 children and the 150 parents, resulting in generations of 300 individuals, and select the top 150 ($\mu$) according to their fitness value to pass to the next generation.

The evolutionary process consists of 20 iterations, thereby generating a total of 3,000 individuals---each representing a distinct hyperparameter configuration. The choice of 20 iterations was determined based on convergence analysis, where we observed that fitness improvements plateaued after approximately 15--20 generations. From this pool of configurations, the one exhibiting the highest fitness is ultimately chosen as the optimal outcome.

\subsection{Generating a default score with BERT}\label{subsec3}

\begin{figure}
    \centering
    \includegraphics[width=0.9\linewidth]{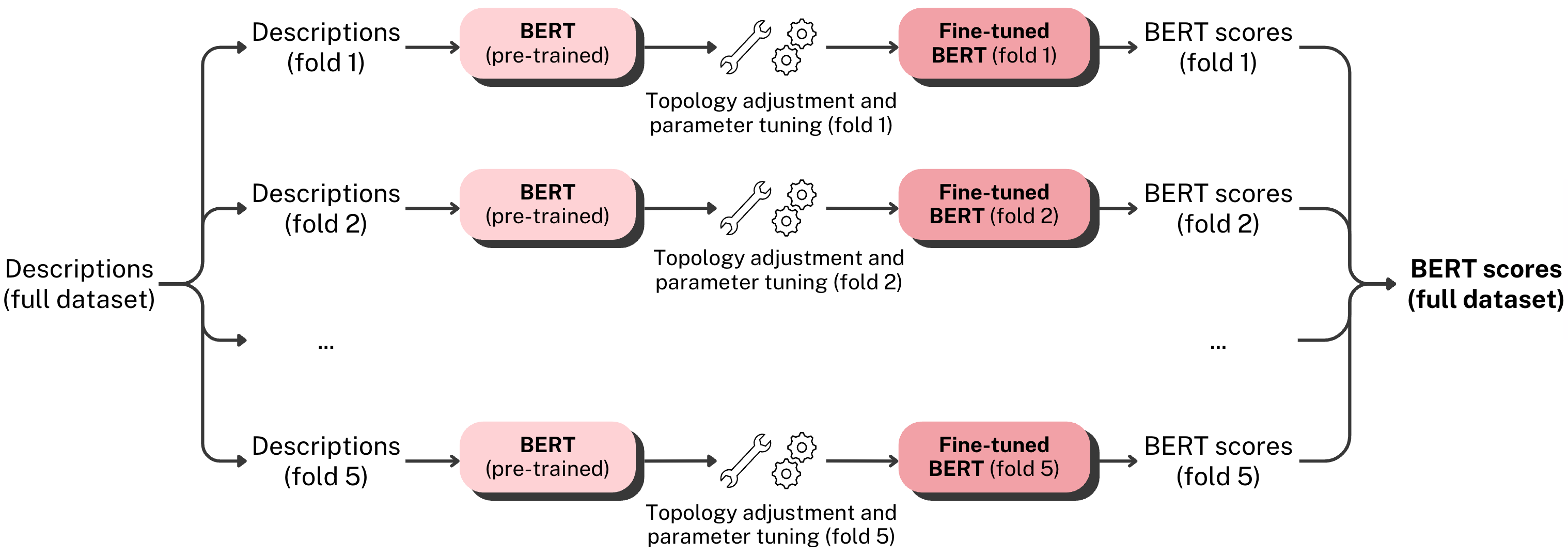}
    
    \caption{Fine-tuning process of BERT for generating a default score from loan descriptions.}
    \label{fig:bert_finetuning}
\end{figure}

In this section, we delineate the methodology employed to generate a default score based on the textual description of the loan. We initiate the process by applying transfer learning utilizing an LLM, specifically BERT in our case. The fine-tuning of BERT, illustrated in Figure \ref{fig:bert_finetuning}, results in a model that produces an outcome within the range of 0 to 1, offering a nuanced indicator rather than a binary classification. This subtle indicator is subsequently integrated into a classifier along with other input variables to predict the likelihood of loan default. The subsequent steps in this process are detailed next.

\subsubsection{Transfer learning to produce the default score}
\label{ssec: Transfer learning}

The BERT model we utilize\footnote{Source: \url{https://tfhub.dev/tensorflow/bert_en_uncased_L-12_H-768_A-12/4}} is configured with L=12 hidden layers (i.e., Transformer encoder blocks), each with a size of H=768, and it employs A=12 attention heads. These attention heads enable the model's self-attention mechanism to process inputs in 12 distinct patterns simultaneously. The output from BERT are embeddings of size 768. For incorporating this model, TensorFlow HUB was selected for its efficient integration with additional neural network layers.

As outlined in Section \ref{sec:BERT}, during the transfer learning phase, we aim to exploit BERT's advanced language understanding while minimizing the need to learn from scratch. To achieve this, we freeze the weights of all but the last hidden layer of BERT. This approach preserves the model's pre-trained capabilities and prevents overfitting on our dataset, while also mitigating catastrophic forgetting---a phenomenon where neural networks lose previously acquired knowledge when retrained on new tasks \cite{biesialska-etal-2020-continual}.

Subsequently, in the fine-tuning stage, only the last BERT layer and the newly added layers are adjusted to better serve our specific task of generating a default score. These layers are expected to enhance the model's adaptability to our particular requirements, allowing slight parameter adjustments for improved task-specific performance, while maintaining the general language understanding gained from BERT's initial pre-training. This strategy effectively balances specialized learning with the retention of valuable pre-trained knowledge.

To ensure that configurations are chosen based on empirical evidence, we incorporate architectural features into the training process. This allows the performance metric---in our case, balanced accuracy---to objectively guide model selection and minimize potential bias from manually imposed configurations. The parameters defining the extra layers' architecture play a crucial role in balancing the model's learning capacity and generalization ability---where, for instance, the second layer enhances learning, and the dropout layer promotes generalization. Thus, we explore various parameter configurations to determine the one that produces the best results. Specifically, we explore the 126 configurations that result from combining the following options:
\begin{itemize}
    \item Using a first extra dense layer of 128, 256, or 512 neurons.
    \item Using or not a second extra dense layer of 128 neurons.
    \item Adding a dropout layer before or after all the extra layers.
    \begin{itemize}
        \item Considering a dropout percentage of 0\%, 0.10\%, 0.20\% or 0.30\% for all the dropout layers.
    \end{itemize}
    \item Using a learning rate of 0.001, 0.0001, or 0.00001.
\end{itemize}

All the internal hidden layers are dense layers using the \textit{ReLU} activation function\footnote{The Rectified Linear Unit (ReLU) activation function outputs the maximum of zero and the input value, ``activating'' the neuron if the input is positive.}. Additionally, to obtain the probability of belonging to the default class, the last layer of the neural network was configured with a single neuron and a sigmoid activation function\footnote{The sigmoid activation function introduces non-linearity and maps the input values to a range between 0 and 1, facilitating binary classification tasks.}.

The neural network is trained to predict the loan outcome (default or non-default) based on the textual description as input. As a loss function, we use the weighted binary cross-entropy, which quantifies the difference between the predicted probabilities and the true binary labels by assigning different weights to each class. This approach ensures that the model does not bias predictions towards the majority class in imbalanced datasets, as it penalizes errors on the minority class more heavily. Considering class weights is crucial given the imbalanced nature of our dataset, where defaulted loans constitute only 15.27\% of the total instances. The training was set up with a batch size of 64 and trained for 25 epochs, with an early stopper of 3 epochs.

The fine-tuning of the BERT model was performed on a system equipped with an Intel Core i9-12900KS processor, an NVIDIA GeForce RTX 4090 graphics card (24GB), and 128 GB of RAM. In contrast, the inference process with the fine-tuned model, which involves generating the BERT score for a given loan description, can be conducted on a standard PC. This capability enhances the practicality of deploying our approach in real-world applications.

\subsubsection{Avoiding data leakage in cross-validation}
\label{ssec: data leakage}

As previously mentioned, our BERT model generates a default probability, which is then integrated into the classifier as an additional quantitative variable. It is important to note that when computing the BERT default probability, we must replicate the exact folds used in the \textit{k}-fold cross-validation of the boosting algorithms. In each iteration, boosting algorithms are trained using the \textit{BERT\_score} variable. It is essential to prevent BERT from training with validation data from a specific fold to avoid distorting the model's true performance. Allowing this would incorporate BERT predictions from its training phase in the fold's validation data, which doesn't accurately represent the model's real-world default prediction ability.

\begin{figure}[]
    \centering
    \includegraphics[width = \textwidth]{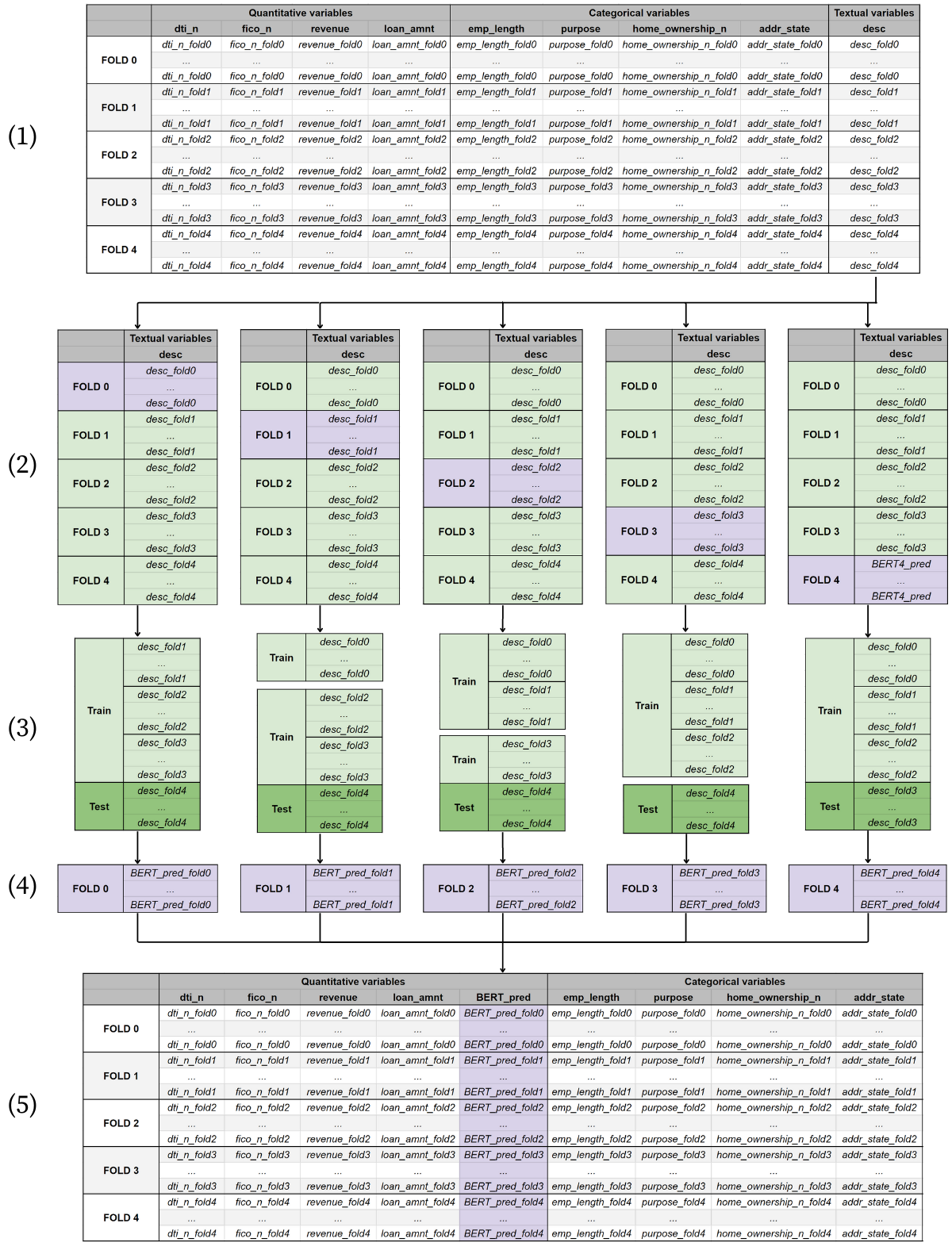}
    \caption{Data handling strategy to prevent data leakage in model training and evaluation.}
    \label{fig:BERT_trainingDiagram}
\end{figure}

To avoid this data-leakage problem, we use the data as shown in Figure \ref{fig:BERT_trainingDiagram}. This diagram consists of five steps:

\begin{enumerate}
    \item Description extraction: Textual descriptions are extracted from the original dataset.
    \item Folds generation: The exact same folds as in the boosting algorithms are generated for the textual descriptions. This is done by dividing the data into a train set (green color) and a test set (purple color). Each fold gets a different test set.
    \item Optimization of the neural network architecture: The train set obtained in the previous step is divided into a 70\% train subset (light green color) and a 30\% test subset (dark green color). The neural network is trained with the previously mentioned configurations on the training subset (light green color), and is tested by predicting the test subset (dark green color).
    \item Default prediction: The optimal configuration obtained in step 3 (lower value of weighted binary cross-entropy in the test subset) is trained with the train and test subsets (light and dark green colors) and is used to predict the default probabilities of the test set obtained in step 2 (purple color). These predictions of unseen data will be the \textit{BERT\_score} values of the current fold.
    \item \textit{BERT\_score} integration: Once the \textit{BERT\_score} values of all the folds are generated in step 4, they are incorporated into the original dataset as a quantitative variable.
\end{enumerate}

\subsubsection{Result of the transfer learning process}
\label{subsubsec3}

As explained before, we explored 126 neural network configurations in each of the 5-fold cross-validation. The resulting optimal configuration for each validation fold and its loss score are shown in Table \ref{tab:BERT_optimalConfig}. Interestingly, the optimal combination of parameters varies across the different folds, and the only parameter that remains constant is the use of the second dense layer. Despite this parameter heterogeneity, the test loss values are quite stable, except for the case of fold 1, which has a slightly higher loss.

\begin{table}[]
    \centering
    \small
    \caption{BERT optimal configuration and score by fold.}\label{tab:BERT_optimalConfig}%
    \begin{adjustbox}{center}
    \begin{tabular}{l|rrrr|r}
        \toprule
             & \begin{tabular}[c]{r}Neurons in\\ 1st dense layer\end{tabular} &
            \begin{tabular}[c]{r}Use 2nd dense layer \\ (128 neurons)*\end{tabular} &
            Dropout &
            Learning rate & 
            Loss** \\
        \midrule
            Fold 0 & 512 & True & 20\% & 0.0001 & 0.1774 \\
            Fold 1 & 512 & True & 0\% & 0.00001 & 0.1793 \\
            Fold 2 & 128 & True & 20\% & 0.0001 & 0.1776 \\
            Fold 3 & 256 & True & 0\% & 0.001 & 0.1775 \\
            Fold 4 & 512 & True & 0\% & 0.001 & 0.1773 \\
        \bottomrule
        \multicolumn{6}{l}{\footnotesize * Values `True' or `False' indicate whether the optimal configuration has or has not that layer.} \\
        \multicolumn{6}{l}{\footnotesize ** Weighted binary cross-entropy of the test set predictions.}
    \end{tabular}
    \end{adjustbox}
\end{table}

\section{Analysis of the BERT Score}
\label{sec:BERTscoreAnalysis}

\subsection{Assessment of the BERT score as a credit risk score}

\begin{table}[h]
    \centering
    \footnotesize
    \caption{Loan descriptions with highest (top) and lowest (bottom) BERT score.}\label{tab:bert_scores}
    \begin{tabularx}{\textwidth}{@{}clX@{}}
        \toprule
        BERT score & Real value & Description \\
        \midrule
        0.8562 & 0 (Non-Default) & \textit{getting a divorce need new apt. with new furniture because she getting everthing.} \\
        0.8149 & 1 (Default) & \textit{need help my bills. to help pay my medications and some bills.}\\
        0.8131 & 0 (Non-Default) & \textit{i can pay of some bills for my self because i been helping other people out. i could save more for my family and their need.i have a good job that i am bless with. i am from a large family seem like every one thinking i suppose to help them when i need help my self.i alway belive that the lord will.}\\
        0.8051 & 0 (Non-Default) & \textit{consolidating our debt makes our life easier live in our means with one solid low monthly payment insted of multiple payment that add up more then what ill be paying with this loan and have a little left to leave in my savings for a rainy day to be honest and thank you for your consideration good dy.}\\
        0.8045 & 0 (Non-Default) & \textit{To consolidate debt. to pay off dept}\\
        
        \midrule
        0.0735 & 0 (Non-Default) & \textit{Debt consolidation with a lower APR.} \\
        0.1146 & 0 (Non-Default) & \textit{In need of funds to pay off some bills as well as minor improvements to house and yard. I have an extremely secure career, and maintaining my credit worthiness is important to me.} \\
        0.1262 & 0 (Non-Default) & \textit{Hard working individual with a stable job will use loan proceeds to consolidate outstanding credit cards balances. 1) Net monthly income - \$4,432. 2) All expenses (allocated):. Rent - \$1,124. Utilities- 84. Groceries 293. Auto (including fuel) 201 . Cell Phone 52. Cable/Internet 64. Personal care items 82. Entertainment/dining 93. Sales tax 65. 3) Previously answered. 4) No.} \\
        0.1360 & 0 (Non-Default) & \textit{This loan is to pay off credit ca.} \\
        0.1871 & 1 (Default) & \textit{I need money for moving expenses and for a buffer for the first month while I transition into working in my new location. I have successfully paid off two previous Lending Club loans in the past couple of years.} \\
        
        \bottomrule
    \end{tabularx}
\end{table}

In this section, we evaluate the effectiveness of the BERT score in utilizing loan descriptions to predict default risk. Table \ref{tab:bert_scores} presents loan descriptions with the highest and lowest BERT scores, which indicate loans assessed by BERT as having the highest and lowest default risks, respectively. It appears that descriptions with higher BERT scores are often less informative, containing errors such as typos (e.g., \textit{``pay of''}, \textit{``everthing''}, \textit{``alway''}, \textit{``belive''}) or grammatical issues. Interestingly, only one of these loans ultimately defaulted. Conversely, loan descriptions with lower BERT scores tend to be more detailed and insightful, often including information about the loan's purpose and the borrower's creditworthiness, and sometimes even providing numeric data related to the borrower's financial status.

\begin{figure}[h]
    \centering
    \includegraphics[width = 0.52\textwidth]{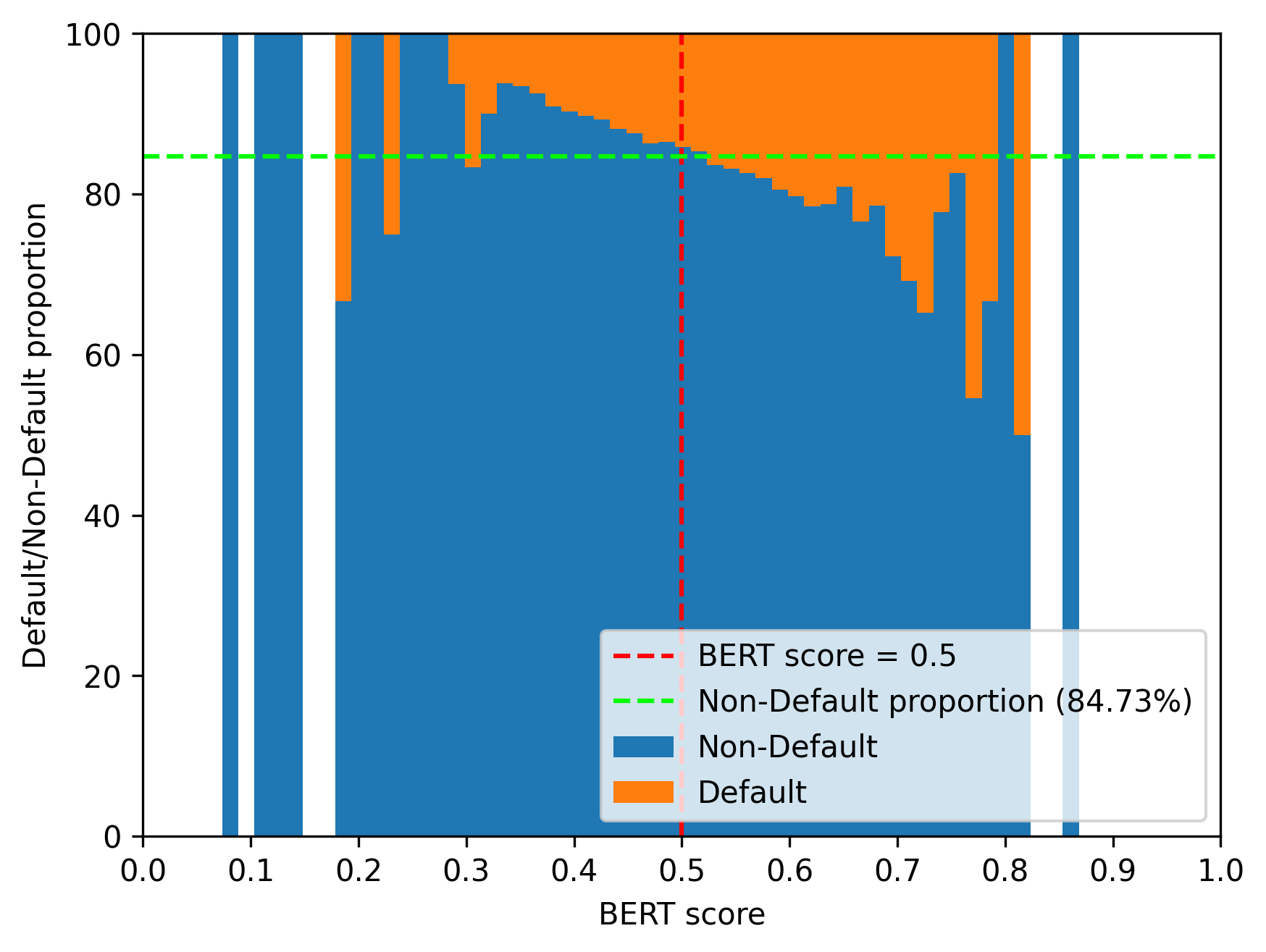}
    \caption{Default and non-default proportions by BERT score range.}
    \label{fig:BERT_bins}
\end{figure}

Figure \ref{fig:BERT_bins} illustrates the distribution of default and non-default loans (\textit{Y} axis) across the BERT score range (\textit{X} axis) in bins of 0.01. In the bar chart, the blue segment represents the percentage of non-default loans, while the orange segment denotes the percentage of default loans\footnote{Absence of a bar indicates that no loan descriptions fall within that BERT score range.}. The figure reveals a general trend where higher BERT scores are associated with a higher proportion of defaulted loans. It is important to note that observations outside the BERT score range of $[0.3, 0.7]$ are sparse, which makes the bars in these regions less reliable. Overall, the trend suggests that higher BERT scores are indicative of a greater likelihood of default, highlighting the BERT score's usefulness as a risk assessment tool.

\begin{table}[]
    \centering
    \small
    \caption{Classification performance of BERT binarization at 0.5.}\label{tab:BERT_binarization}%
    \begin{tabular}{@{}lccccccc@{}}
        \toprule
            & & \multicolumn{3}{c}{Default} & \multicolumn{3}{c}{Non-default} \\
            \cmidrule(lr){3-5} \cmidrule(lr){6-8}
            Model & BACC & Precision & Recall & F1 & Precision & Recall & F1\\
        \midrule
            BERT & 0.5444 & 0.1714 & 0.6896 & 0.2746 & 0.8771 & 0.3993 & 0.5487 \\
        \bottomrule
    \end{tabular}
\end{table}

Table \ref{tab:BERT_binarization} shows the classification performance obtained by applying a 0.5 threshold to binarize the BERT score, which obtains a balanced accuracy of 54.4\%. The BERT score is not very precise in predicting the default class (17.1\%) but retrieves 69\% of the instances.

These results demonstrate that the BERT model can be effectively fine-tuned to predict the final state of the loan using only the information provided by the borrower in the description field of the application form. In Section \ref{sec: Classification Results}, we will contextualize these findings by comparing them with the results obtained from XGBoost and various sets of variables.

\subsection{Relationship between the BERT score and other variables}
\label{ssec:BERTscore_and_other_vars}
We now explore potential relationships between the BERT score and other variables. Table \ref{tab:BERT_corr_num} presents the correlation coefficients between the BERT score and the quantitative variables within the dataset. The findings reveal weak yet statistically significant relationships with all variables except for the loan amount. Notably, the most pronounced correlations exist with the FICO score and revenue variables. Both exhibit inverse relationships, indicating that individuals with higher FICO scores and revenues tend to have lower BERT scores, and vice versa. The association between the BERT score and the debt variable (\textit{dti\_n}) is direct, albeit slightly weaker than the other two correlations.

\begin{table}[]
    \centering
    \small
    \caption{Correlation coefficients of quantitative variables with the BERT score.}\label{tab:BERT_corr_num}
    \begin{tabular}{lrr}
        \toprule
            Variable & Pearson & Spearman \\ 
        \midrule
            revenue & -0.0734* & -0.1007* \\
            dti\_n & 0.0663* & 0.0627* \\
            loan\_amnt & -0.0008 & -0.0012 \\
            fico\_n & -0.1293* & -0.1293* \\ 
        \bottomrule
        \multicolumn{3}{l}{\footnotesize * p-value less than 0.01.}
    \end{tabular}
\end{table}

\begin{table}[]
    \centering
    \small
    \caption{Kruskal-Wallis test of categorical variables with the BERT score.}\label{tab:BERT_kw}
    \begin{tabular}{lr}
        \toprule
            Variable & H-statistic \\
        \midrule
            emp\_length & 1581.67* \\
            purpose & 1357.94* \\
            home\_ownership & 243.20*\\
            addr\_state & 360.44*\\
        \bottomrule
        \multicolumn{2}{l}{\footnotesize * p-value less than 0.01.}
    \end{tabular}
\end{table}

To evaluate the relationship with the categorical variables, Table \ref{tab:BERT_kw} presents the results of the Kruskal-Wallis test, a non-parametric statistical test that analyzes whether there are differences in the BERT scores across categories within each categorical variable. The results indicate significant differences in BERT scores among all categorical variables, suggesting a certain level of association between the BERT score and these categorical factors. However, it remains challenging to quantify the strength of this relationship or identify the specific categories with the most robust associations.

\begin{table}[]
    \centering
    \small
    \caption{Correlation coefficients of linguistic features with the BERT score.}
    \label{tab:BERT_corr_ling}
    \begin{tabular}{lrr}
        \toprule
            Variable & Pearson & Spearman \\ 
        \midrule
            Word count & -0.1961* & -0.2650* \\
            Polarity & -0.1036* & -0.1473* \\
            Subjectivity & -0.1753* & -0.1866* \\
            Readability & 0.1182* & 0.1518* \\ 
        \bottomrule
        \multicolumn{3}{l}{\footnotesize * p-value less than 0.01.}
    \end{tabular}
\end{table}

Finally, Table \ref{tab:BERT_corr_ling} examines the potential relationship between the BERT score and various linguistic features automatically extracted from the text. We find statistically significant correlations between the BERT score and all linguistic features analyzed. Notably, there is a strong inverse correlation with both word count and subjectivity, indicating that shorter and more objective texts tend to have higher BERT scores. This finding, together with the correlations presented in Table \ref{tab:BERT_corr_num}, suggests that BERT is more closely related to linguistic features than to the numerical variables associated with loan applications.

\section{Results of the LLM-Enhanced Granting Model}
\label{sec: Classification Results}

\subsection{Analysis of the classification performance}
First, we evaluate the impact of incorporating the BERT score in the granting model. In our baseline experiment, we optimize XGBoost with a genetic algorithm using the quantitative and categorical variables typically used in granting models, while in the competing approach, we optimize it but include the BERT score as an input variable. Table \ref{tab:results} shows the results of both approaches.

A closer examination of the balanced metrics reveals a marginal enhancement in both BACC (0.6154 vs. 0.6187) and AUC (0.6575 vs. 0.6644), the latter significant according to the DeLong test \cite{DeLongTest}---a non-parametric approach for evaluating whether differences between AUCs of two models are statistically significant--- at the 0.01 level. It is crucial to note that the Lending Club dataset exclusively consists of approved loans. This fact poses a substantial challenge to significantly enhance the outcomes in a loan granting model such as ours since the loans included in the dataset were initially considered favorable by the platform. Furthermore, in experiments not reported in the paper we used CatBoost \cite{CatBoost2018} instead of XGBoost and obtained a similar BACC improvement.

\begin{table}[]
    \centering
    \small
    \caption{Classifier performance with and without the BERT score.}
    \label{tab:results}
    \centering
    \begin{tabular}{lcc}
        \toprule
            Metric & Quant. + Categ. var. & Quant. + Categ. + BERT score \\
        \midrule
            BACC & 0.6154 & 0.6187 \\
            AUC & 0.6575 & 0.6644 \\
            F1 & 0.3266 & 0.3308 \\
            Precision & 0.2168 & 0.2249 \\
            Recall & 0.6614 & 0.6360 \\
            Accuracy & 0.5835 & 0.6066 \\
        \bottomrule
    \end{tabular}
\end{table}

The performance metrics in Table \ref{tab:results} indicate that incorporating the BERT score results in improved precision but diminished recall. However, attributing this change in precision-recall behavior solely to the BERT score requires careful consideration. This caution arises from our observations within our dataset, where classifiers with different hyperparameters and similar near-optimal balanced accuracy values have demonstrated varying precision-recall behaviors, suggesting that this may also be the case here.

Table \ref{tab:results_individual} shows an additional experiment in which XGBoost classifiers are trained and optimized using only one kind of input variable. While the classifier using the quantitative variables is clearly the best, the classifier that uses just the BERT score obtains slightly better results than the one using the four categorical variables (the AUC difference is significant at 0.01  according to the DeLong test). This is noteworthy given the well-known effectiveness and the meaningful nature of the qualitative variables. 

Table \ref{tab:results_individual} also shows the result of an XGBoost classifier using the textual features presented in Table \ref{tab:BERT_corr_ling}, namely: polarity, subjectivity, word count, and readability score. This classifier is outperformed by the XGBoost that uses the BERT score in terms of balanced accuracy and AUC (significant at the 0.01 level according to the DeLong test). This finding underscores the superior ability of the fine-tuned LLM to leverage textual descriptions and extract relevant information for the classification task.

\begin{table}[]
    \small
    \caption{Performance of the XGBoost considering only a subset of variables.}
    \label{tab:results_individual}
    \centering
    \begin{tabular}{lcccc}
        \toprule
            Metric & Quant. & Categ. & Text. & BERT score \\
        \midrule
            BACC & 0.6062 & 0.5486 & 0.5258 & 0.5490 \\
            AUC & 0.6457 & 0.5656 & 0.5309 & 0.5714 \\
            F1 & 0.3192 & 0.2759 & 0.2534 & 0.2601 \\
            Precision & 0.2138 & 0.1746 & 0.1665 & 0.1877 \\
            Recall & 0.6300 & 0.6563 & 0.5302 & 0.5153 \\
            Accuracy & 0.5896 & 0.4738 & 0.5227 & 0.5724 \\
        \bottomrule
    \end{tabular}
\end{table}

\subsection{Feature importance and explainability}

\begin{figure}[]
    \centering
    \begin{subfigure}{0.5\textwidth}
        \centering
        \includegraphics[width=\linewidth]{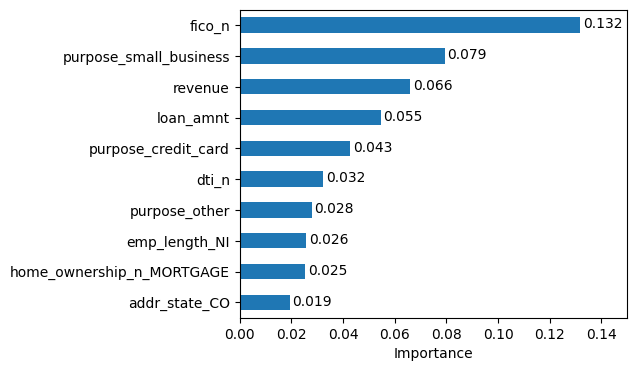}
        \caption{Before BERT score incorporation.}
        \label{fig:feat_imp_noBERT}
    \end{subfigure}%
    \begin{subfigure}{0.5\textwidth}
        \centering
        \includegraphics[width=\linewidth]{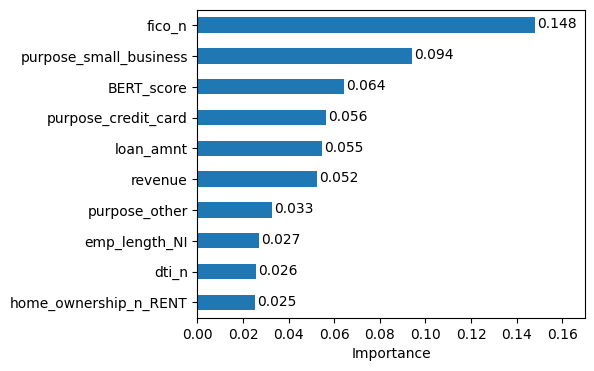}
        \caption{After BERT score incorporation.}
        \label{fig:feat_imp_BERT}
    \end{subfigure}

    \caption{Feature importance of the two XGBoost classifiers.}
    \label{fig:feat_imp}
\end{figure}

Now we delve into how the inclusion of the BERT score alters the classifier's use of input variables. Our goal is not only to assess the predictive contribution of the BERT score but also to understand its broader impact on the model's decision-making process. Figure \ref{fig:feat_imp} shows the feature importance assigned by XGBoost, both with and without the BERT score. Notably, the BERT score becomes the third most influential variable, accounting for 6\% of the total importance. Moreover, its inclusion reshapes the importance and ranking of other features. For instance, the importance of the borrower's annual revenue and the debt-to-income ratio (\textit{dti\_n}), both crucial for assessing the borrower's economic status, significantly decreases. 
This shift suggests that the BERT score may encapsulate information that overlaps with these variables or renders them less critical, as supported by the correlation analysis in Table \ref{tab:BERT_corr_num}.

\begin{figure}[]
    \centering
    \begin{subfigure}{0.5\textwidth}
        \centering
        \includegraphics[width=\linewidth]{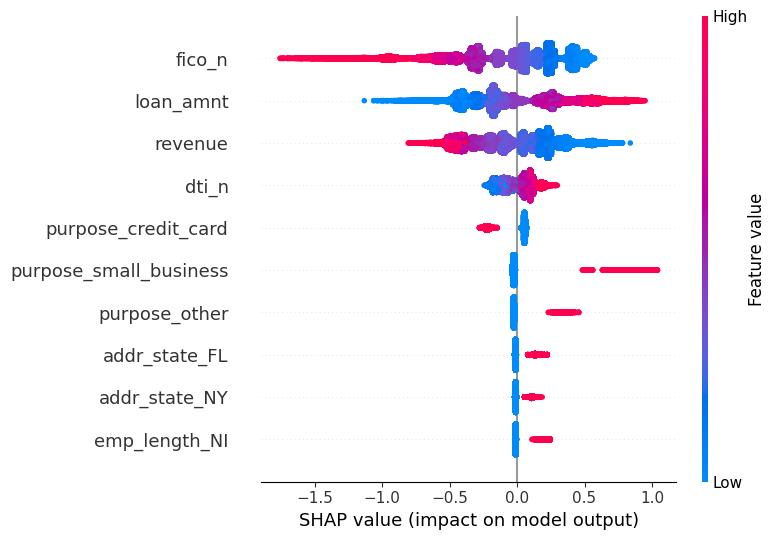}
        \caption{Before BERT score incorporation.}
        \label{fig:SHAP_noBERT}
    \end{subfigure}%
    \begin{subfigure}{0.5\textwidth}
        \centering
        \includegraphics[width=\linewidth]{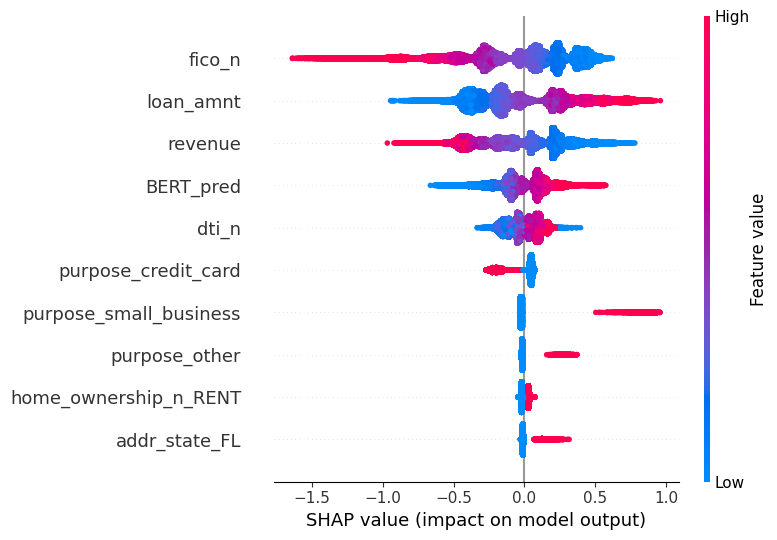}
        \caption{After BERT score incorporation.}
        \label{fig:SHAP_BERT}
    \end{subfigure}
    \caption{SHAP values of the XGBoost classifiers.}
    \label{fig:SHAP}
\end{figure}

To analyze whether the relationships between each variable and the predicted outcome change, we use the SHAP values \cite{LundbergSHAP}, which quantify the contribution of each feature to the model's predictions. Figure \ref{fig:SHAP} compares the SHAP values for the 10 most impactful features in models both with and without the BERT score. It reveals that the distributions of SHAP values, as seen in the violin plots, do not significantly change when the BERT score is included. Interestingly, the SHAP value distribution in Figure \ref{fig:SHAP_noBERT} mirrors a similar analysis presented in \cite{ArizaExplainability}, conducted on a lending model with Lending Club data\footnote{Our dataset is narrower, considering only loans accompanied by textual descriptions, as detailed in Section \ref{sec:Dataset_des}.}, which aligns with expectations in the credit risk context.

\begin{figure}[h]
    \centering
    \includegraphics[width=0.5\linewidth]{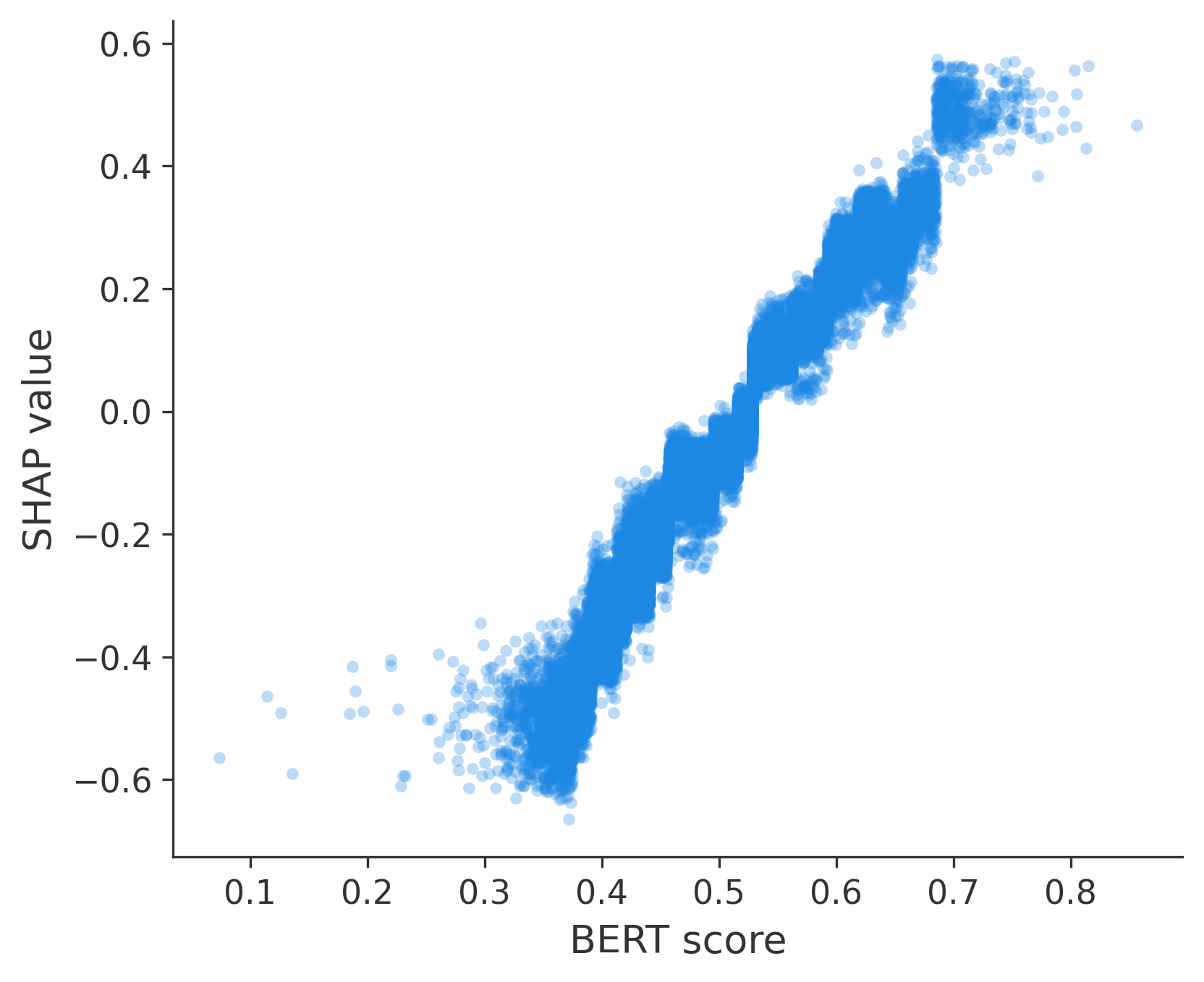}
    \caption{Dependence plot between SHAP values and BERT score.}
    \label{fig:SHAP_BERT_score}
\end{figure}

Figure \ref{fig:SHAP_BERT_score} reveals a direct and linear relationship between the BERT score and the SHAP values, which in this case relate to the default risk. Notably, this relationship is asymmetric around BERT score values of 0.5; scores below 0.4 correspond to SHAP values ranging between -0.4 and -0.6, strongly guiding the model toward predicting non-default. Conversely, this impact range in the positive case is only reached by BERT scores exceeding 0.7, signifying that only exceptionally high BERT scores serve as strong indicators of default.

\subsection{A closer examination of the role of the BERT score in classification}

\begin{figure}[]
    \centering
    \begin{subfigure}{0.5\textwidth}
        \centering
        \includegraphics[width=\linewidth]{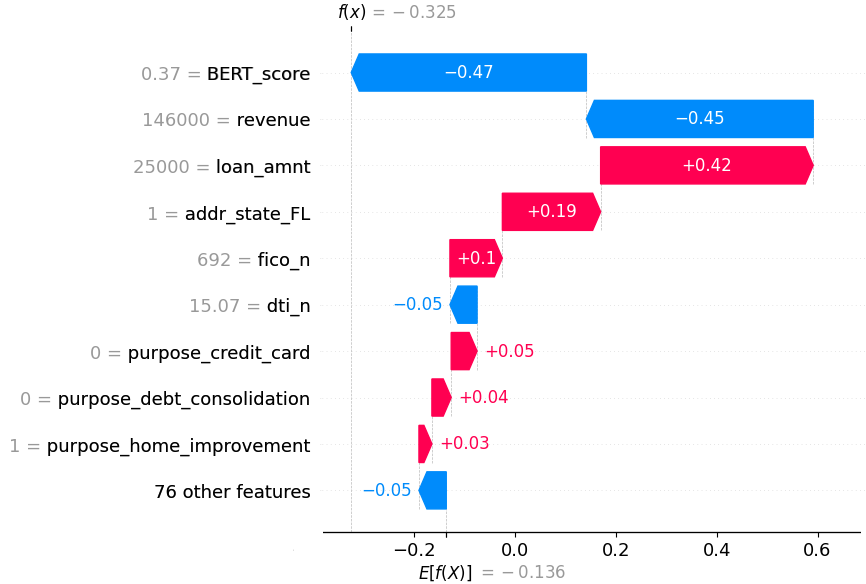}
        \caption{Loan (correctly) classified as non-default.}
        \label{fig:waterfall_NONDEF}
    \end{subfigure}%
    \begin{subfigure}{0.5\textwidth}
        \centering
        \includegraphics[width=\linewidth]{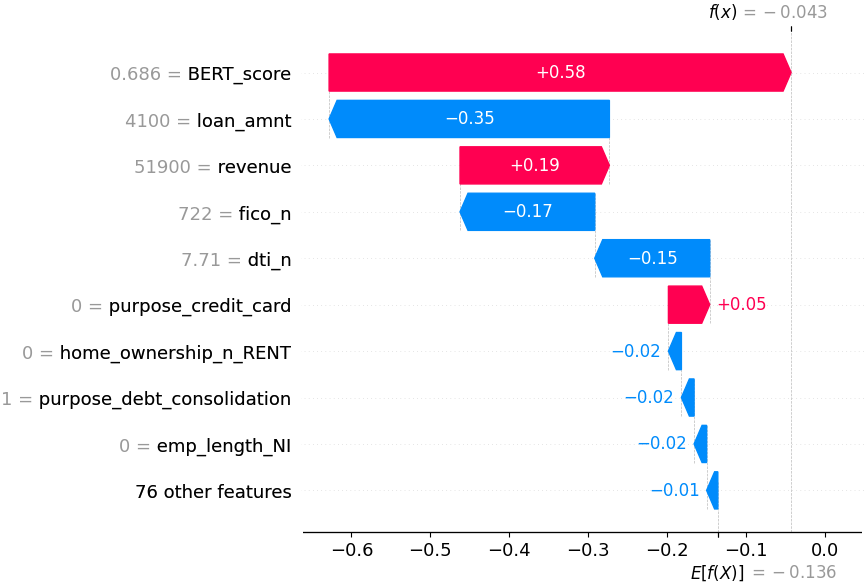}
        \caption{Loan (correctly) classified as default.}
        \label{fig:waterfall_DEF}
    \end{subfigure}

    \caption{Waterfall plots representing the SHAP values of the features in two loans.}
    \label{fig:waterfall}
\end{figure}

As shown in our analysis of feature importance in XGBoost, the BERT score plays a crucial role in the correct classification of the loans. Using SHAP values, we can determine how each variable influences the classification decision for a given loan. Figure \ref{fig:waterfall} shows the waterfall plot for two loans where the BERT score is a decisive factor. The plot should be read from bottom to top. It begins at the bottom with the expected value of the model output. Each row then shows how each feature's positive (red) or negative (blue) contribution shifts the prediction from this expected value to the final prediction for that specific loan. Features are ordered from bottom to top in ascending order of contribution (in absolute value). The x-axis represents log-odds units (the margin output before the logistic link function used by the classifier), meaning that negative values correspond to probabilities below 0.5 for classifying the loan as default.

Figure \ref{fig:waterfall_NONDEF} shows a loan with a low BERT score (0.37) and a SHAP value of $-0.47$, the highest in absolute value for that instance. This SHAP value counteracts the influence of other variables, leading to a (correct) classification of the loan as non-default. Conversely, Figure \ref{fig:waterfall_DEF} illustrates the opposite scenario: a loan with a high BERT score (0.686), where the SHAP value---again, the highest for that instance---overrides the effect of other variables and correctly assigns the loan to the default class. Below are the description of both loans given by its borrowers:
\begin{quote}
    \small
    \textbf{Purpose: Home improvement; BERT score: 0.37} \\
    \textit{``I want to have an in ground pool built with an aluminum screened enclosure. Nothing super fancy, just enough to enjoy with my wife a kids.''}\\ \\ 
    \textbf{Purpose: Debt consolidation; BERT score: 0.686} \\
    \textit{``I have problems with one of my loans what I have it is with household-beneficial finance bank it is for \$15,250.00 and the interest rate is 25.8\% and the payment is 323 monthly and I have 3 months pass due, I page all time in time but the situacion make loss that way cause the gas bill was very higher and the last mont the second day my Mother die and was very expensive for us her funeral cause she die here but we bring her body to Nicaragua, and the rest is for a loan with American General Finance \$2,500.00, Wells Fargo \$480.00 , Fifth Third Bank Optimun \$357.00, all make a total of \$18,587.00, please I need cause I get some problems with the big loan with household-beneficial.''}\\ 
\end{quote}

Although Section \ref{ssec:BERTscore_and_other_vars} suggested that shorter texts tend to have higher BERT scores, this is not the case here. BERT's ability to analyze content and textual characteristics enables it to correctly associate the long text with a high probability of default and the shorter one with a higher likelihood of repayment.

\subsection{Impact of the BERT score in the classification results across the purpose categories}

In this section, we examine how incorporating the BERT score in the model changed classification outcomes across various categories of the purpose variable. Given the language comprehension capabilities of LLMs, it is conceivable that the BERT score provides a more nuanced characterization of the risk associated to the loan purposes than the categorical \textit{purpose} variable alone. For instance, the inherently ambiguous `other' category may benefit from the nuanced understanding of loan descriptions provided by the BERT model, potentially leading to improved prediction outcomes.

\begin{figure}[]
    \centering
    \includegraphics[width = 0.79\textwidth]{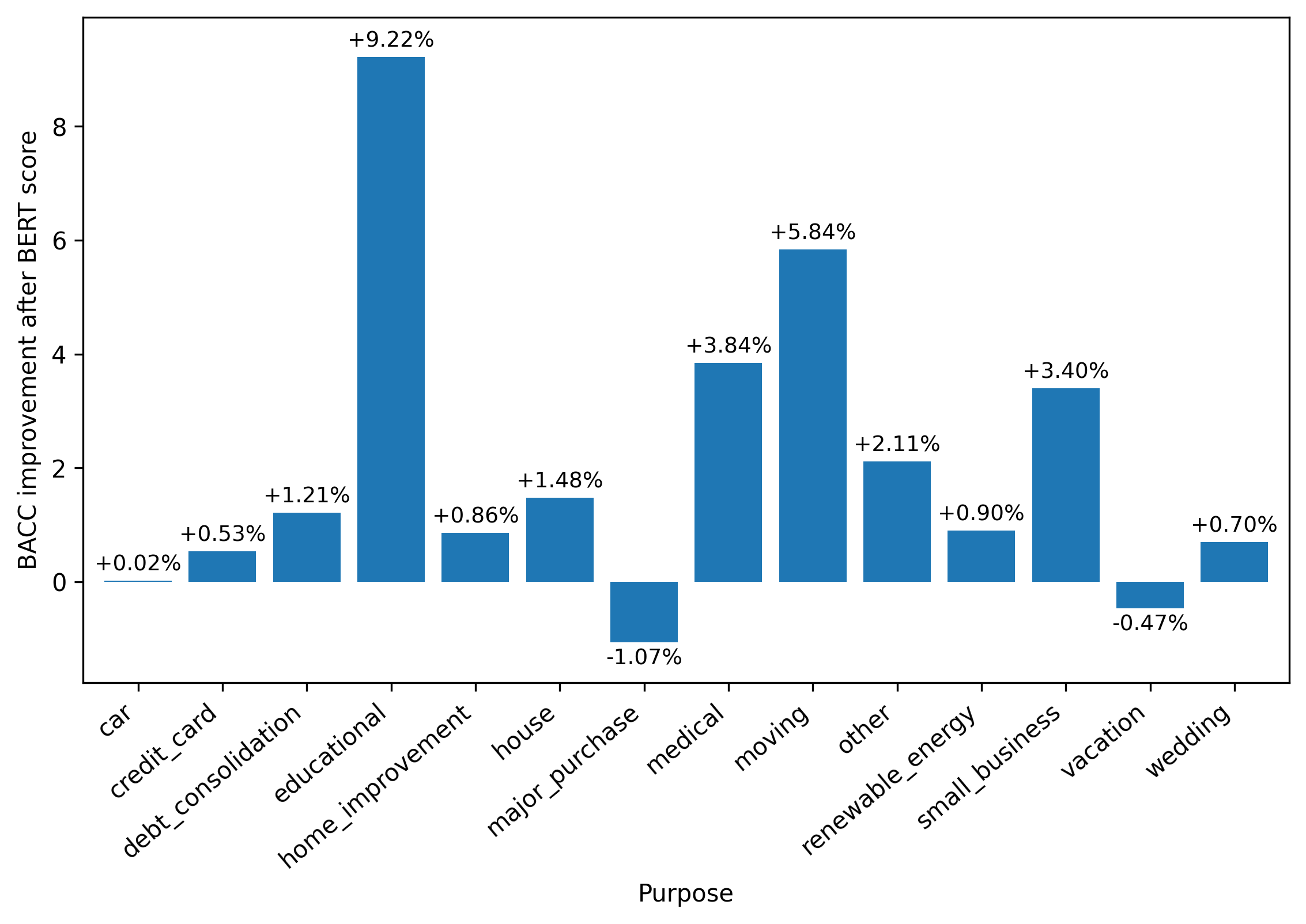}
    \caption{BACC changes (in relative percentage) by purpose after including the BERT score.}
    \label{fig:purpose_improvement}
\end{figure}

Figure \ref{fig:purpose_improvement} illustrates the relative changes in balanced accuracy for each loan purpose after adding the BERT score. Notably, while the `other' category shows a modest improvement of 2.11\%, several other categories exhibit more significant enhancements, including `educational' (9.22\%), `moving' (5.84\%), `medical' (3.84\%), and `small business' (3.40\%). Given the black-box nature of our model, it is not easy to ascertain why these categories have improved more than the others. However, we posit that these categories share a commonality---the more detailed specification of purpose or a deeper understanding of the borrower's situation contributes to a more precise delineation of the default risk. For instance, `education' loans might incorporate information about the borrower's field of study or educational institution, which could correlate with employability and repayment capacity. Similarly, in categories like `moving', `medical', or `small business' loans, the BERT score likely reflects a deeper understanding of the borrower's situation, enabling a more accurate default risk assessment.

The following examples illustrate instances of loans that were initially misclassified as defaults without the BERT score but were correctly predicted as non-defaults after its inclusion:

\begin{quote}
    \small
    \textbf{Purpose: Educational; BERT score: 0.3765} \\
    \textit{``I'm 25 years old and living in New Orleans. I'm asking for a relatively small amount of money to help me take care of my post-Bac tuition for teacher certification and to help me pay off a credit card. I currently work as a private school teacher making very little money with no benefits (about \$29,000 a year). I have to pay about \$1000 in the coming year for my tuition, and I have to get health insurance ASAP, but it's hard to do so with no financial help from anyone else. My parents can't help me because my mother is permanently disabled and my father took a huge pay cut this year.''}\\ \\ 
    \textbf{Purpose: Moving; BERT score: 0.3843} \\ 
    \textit{``Although I can afford payments,due to some recent expenses, I am short on cash flow for an unexpected move. I am, however, looking for a more reasonable alternative to banking rates. I have borrowed from Lending Club before and always paid fully and on time with automatic payments.''}\\ \\ 
    \textbf{Purpose: Small business; BERT score: 0.3292} \\
    \textit{``The purpose of this loan is to fund advertising costs for a growing internet business venture. I am a successful sales professional earning an average of over \$250K per year over the last 5 years. My credit scores are strong and I have a documented history of paying all my debts (personal or business related) on time.''}\\ \\ 
    \textbf{Purpose: Medical; BERT score: 0.3921} \\
    \textit{``This loan will be used to pay off a Care Credit credit card currently at 21.9\% I used the card to pay for a prosthetic limb that my health insurer would not cover.''}\\
\end{quote}

In these cases, the corresponding BERT scores are consistently below 0.4. As shown in Figure \ref{fig:SHAP_BERT_score}, scores below this threshold are strong drivers for predicting non-default. All of these texts offer precise descriptions of the purpose of the loan or the borrower's situation, which allows a relatively moderate risk to be anticipated.

\begin{figure}[]
    \centering
    \begin{subfigure}{0.5\textwidth}
        \centering
        \includegraphics[width=\linewidth]{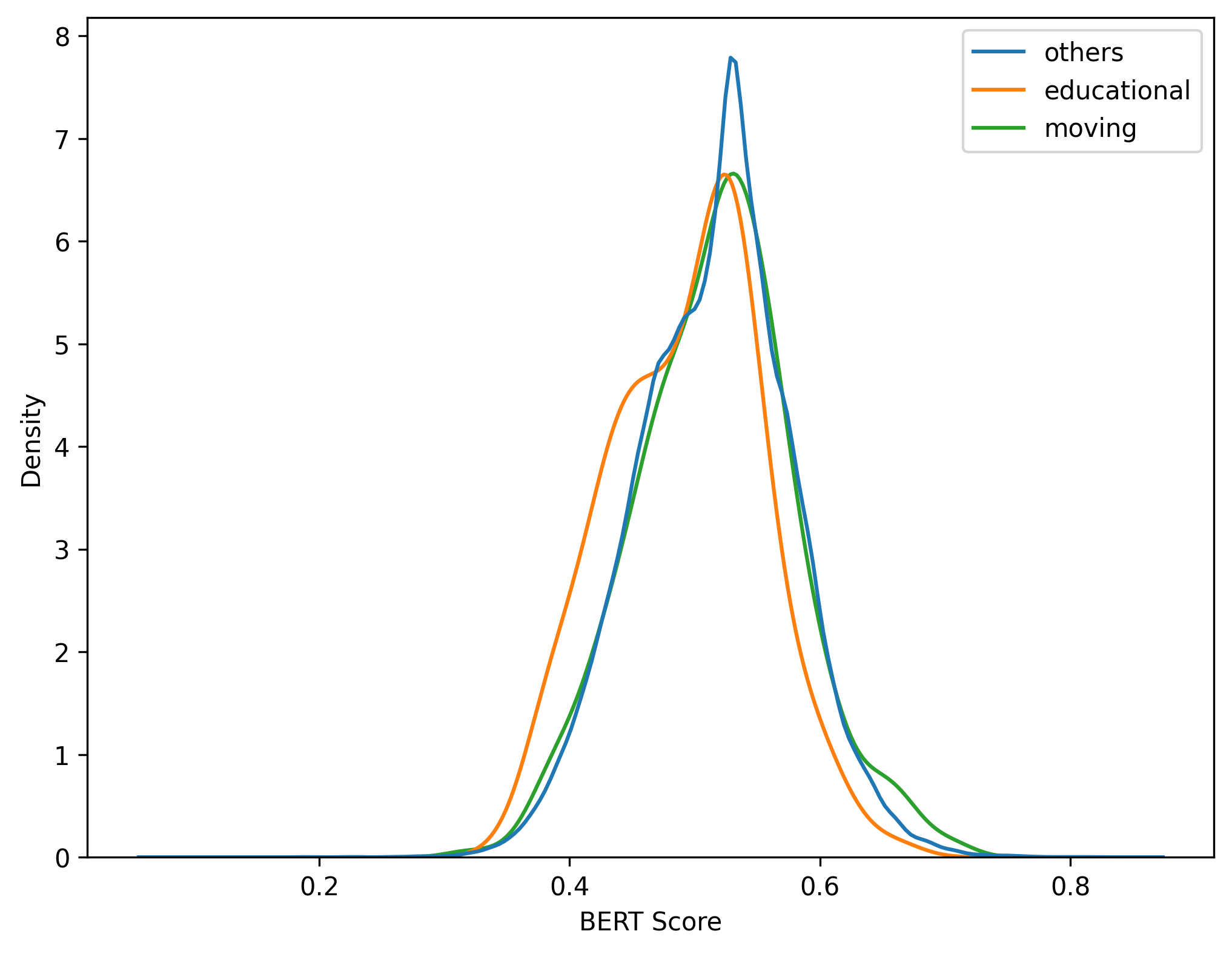}
        \caption{Distributions of the BERT score.}
        \label{fig:BERT_score_fdens}
    \end{subfigure}%
    \begin{subfigure}{0.5\textwidth}
        \centering
        \includegraphics[width=\linewidth]{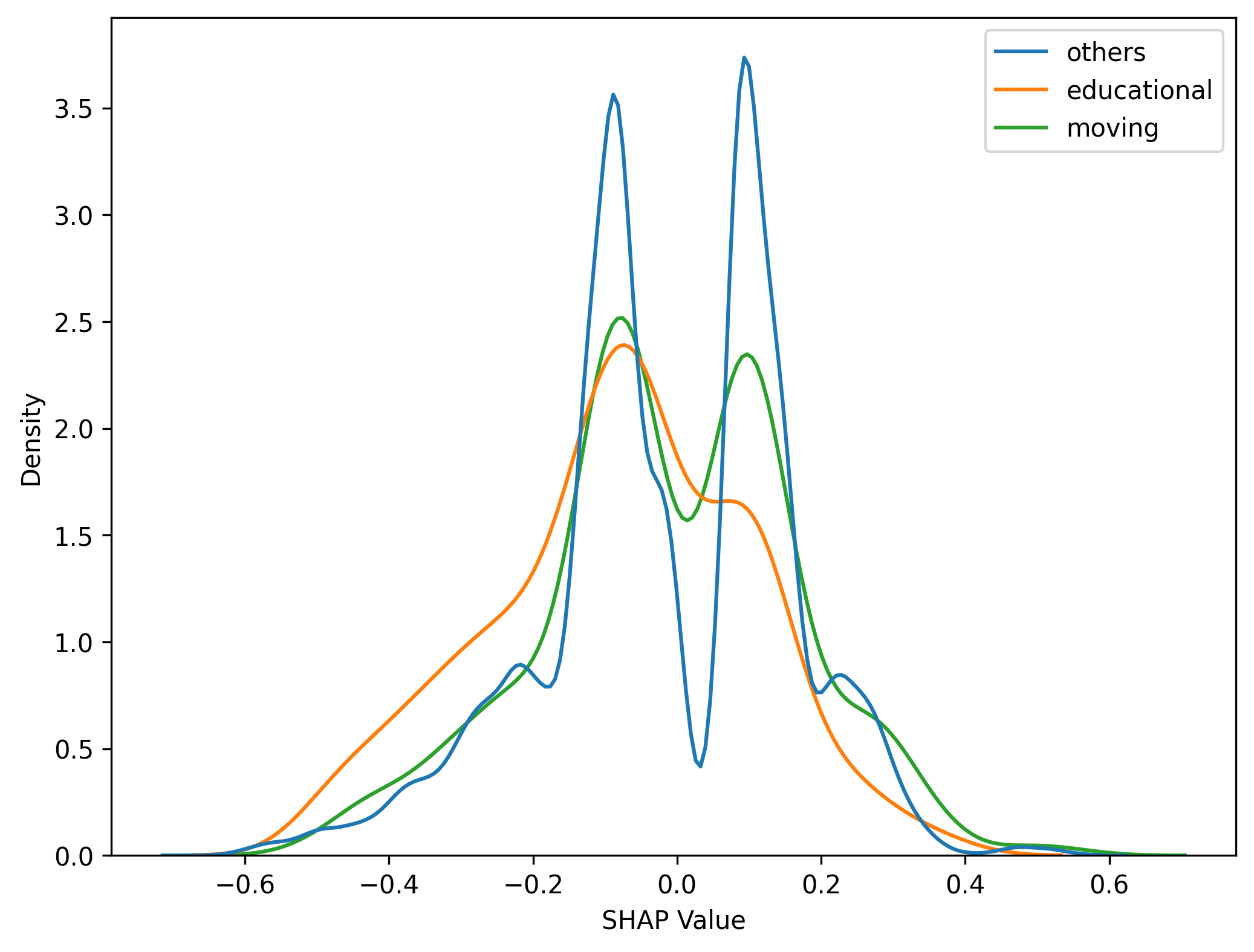}
        \caption{Distributions of the SHAP values.}
        \label{fig:SHAP_fdens}
    \end{subfigure}

    \caption{BERT score and SHAP values distributions for the educational (orange), moving (green), and the rest of purposes (blue).}
    \label{fig:BERT_score_densities}
\end{figure}

To better understand the differences in classifier performance across purposes, we conduct an additional analysis focusing on the purposes with the greatest performance improvement. Figure \ref{fig:BERT_score_fdens} shows the distributions of BERT score values for loans with \textit{educational} and \textit{moving} purposes, compared to the distribution of loans with all other purposes, which serves as the reference. We can see that the reference distribution is more sharply peaked, indicating that most loans have average BERT score values that contribute little information to the classification. In contrast, the distribution of the loans with an \textit{educational} purpose is shifted to the left and exhibits a higher density of low BERT score values (below 0.4), which indicates that the loan will likely be repaid. In the case of the \textit{moving} purpose, the most pronounced difference can be seen in the right tail, as it has a higher frequency of values above 0.6, which helps to detect loans likely to default.

\begin{table}[]
    \small
    \caption{Coefficients of the quantile regression at the 5th, 10th, 50th, 90th, and 95th percentiles.}
    \label{tab:quant_reg_extended}
    \centering
    \begin{tabular}{lrrrrr}
        \toprule
             & 5th & 10th & 50th & 90th & 95th \\
        \midrule
            Intercept & 0.4101 & 0.4328 & 0.5178 & 0.5890 & 0.6106 \\
            \cmidrule{2-6}
            Default & +0.0164* & +0.0166* & +0.0127* & +0.0124* & +0.0146* \\
            Educational & $-$0.0208* & $-$0.0240* & $-$0.0205* & $-$0.0308* & $-$0.0316* \\
            Moving & $-$0.0022 & $-$0.0014 & +0.0002 & $-$0.0027 & +2.8e-05 \\
            \cmidrule{2-6}
            Default--Educational & $-$0.0067 & $-$0.0077 & +0.0126 & +0.0288 & +0.0084 \\
            Default--Moving & $-$0.0085 & $-$0.0040 & +0.0062 & +0.0286* & +0.0357* \\
        \bottomrule
        \multicolumn{6}{l}{\footnotesize * p-value less than 0.01.}
    \end{tabular}
\end{table}

To assess the significance of these differences, we perform a quantile regression with interactions at five percentiles: the median (50th) and the two tails of the distribution (5th, 10th, 90th and 95th). The quantile regression model is specified as:
\begin{equation}
    \text{BERT\_score}_i = \beta_0 
    + \beta_1 \cdot \text{Default}_i 
    + \beta_2 \cdot D_E 
    + \beta_3 \cdot D_M 
    + \beta_4 \cdot (\text{Default}_i \times D_E) 
    + \beta_5 \cdot (\text{Default}_i \times D_M) 
    + \epsilon_i
\end{equation}
Here, $D_E$ and $D_M$ represent the dummy variables for \textit{educational} and \textit{moving} loans, respectively, with all other purposes serving as the reference group. Meanwhile, $\text{Default}_i$ is an indicator variable that denotes whether the loan eventually defaulted or not. The quantile regression allows us to assess whether there are significant differences in the BERT score between defaulted and non-defaulted moving and educational loans compared to loans of other purposes.

The results are presented in Table \ref{tab:quant_reg_extended}. At the 5th, 10th, 50th, 90th and 95th percentiles, the intercept values (0.4101, 0.4328, 0.5178, 0.5890, and 0.6106) capture the baseline BERT score for loans that do not belong to the educational or moving purposes and that were returned (i.e., $D_E=0$, $D_M=0$, and $\text{Default}_i = 0$). The positive and statistically significant coefficients for \textit{default} across all quantiles indicate that default is associated with an increase in BERT score (+0.0164, +0.0166, +0.0127, +0.0124, and +0.0146). Thus, the first conclusion that we can draw is that our fine-tuned BERT successfully assigns a higher risk to the loans that eventually defaulted, at least in the quantiles analyzed, in line with the results obtained in previous analyses.

The \textit{educational} dummy variable shows significant negative coefficients across all quantiles ($-$0.0208, $-$0.0240, $-$0.0205, $-$0.0308, and $-$0.0316), confirming that, regardless of default status, educational loans have lower BERT scores at the quantiles analyzed compared to the reference group---consistent with a lower risk profile observed in Figure \ref{fig:BERT_score_fdens}. Regarding the interaction term, the \textit{Default--Educational} coefficient is not significant at any quantile, suggesting that defaulted educational loans do not have significantly different BERT scores compared to repaid educational loans.

In contrast, the coefficients of the \textit{moving} dummy when the \textit{default} variable is zero are not statistically significant at any quantile. This implies that, in such cases, non-defaulted moving loans do not differ from the non-defaulted reference group in their baseline BERT scores. As for variable interaction, the \textit{Default--Moving} coefficient is significant at the 90th and 95th percentiles (+0.0286 and +0.0357). This result indicates that, at the upper end of the BERT score distribution, defaulted loans from the moving purpose have higher BERT scores than defaulted loans from the rest of the purposes. In other words, the difference observed in Figure \ref{fig:BERT_score_fdens} in the right tail of the BERT score distribution for moving loans relative to the reference group is driven by the higher risk associated with defaulted loans.

Overall, these findings evidence a nuanced relationship: default increases BERT scores across the board, non-defaulted educational loans exhibit lower scores---reflecting lower risk---than other categories, and moving loans that eventually defaulted exhibit a higher risk reflected only in the upper quantiles. To conclude the analysis, Figure \ref{fig:SHAP_fdens} shows the distribution of SHAP values for \textit{educational} and \textit{moving} purposes compared to the rest of the purposes. Again, educational loans show a distribution shifted to the left and with a denser left tail, indicating that XGBoost uses this variable to identify loans likely to be repaid. In the case of moving loans, differences are observed at the extremes, especially in the right tail, indicating the improvement in discrimination that this variable provides to the classifier. This analysis shows that the interaction between the BERT score and the educational and moving purposes is the reason behind the remarkable performance improvements observed in these categories.

As a conclusion, while previous research had reported mixed results regarding the predictive power of linguistic factors for loan default \cite{GaoLin2015_lemon, Wang2016, DORFLEITNER2016, Siering2023}, our findings suggest that an LLM-based risk indicator used at the granting stage has a significant positive influence in the classification results, thus demonstrating BERT's capacity to draw meaningful information from the loan descriptions.

\section{Conclusion}
\label{sec: Conclusion}

In this paper, we presented a novel approach that leverages state-of-the-art natural language processing techniques to enhance credit risk models. By fine-tuning BERT on loan descriptions, we generated a risk score that effectively distinguishes between defaulted and non-defaulted loans, particularly at the granting stage, when decision-makers have limited variables available to inform their decisions. In addition, integrating this BERT-based risk score with traditional variables significantly improved the performance of conventional loan-granting models. This result aligns with those obtained by Xia et al. \cite{Xia2023}.
Our analysis suggests that the information extracted by the language model can capture aspects of the text related to the linguistic aspects but also with content-based factors related to loan purpose and the borrower's creditworthiness. The inclusion of the BERT-based risk score also reshapes the classifier’s decision-making process and the role played by other variables.

Our approach can be easily applied without the need for manual annotation, which is a complex and subjective task. Additionally, while fine-tuning the LLM is a computationally intensive process that requires GPU resources, generating predictions such as the risk score for a loan description can be done rapidly on standard home equipment, making this approach highly accessible. This work opens several avenues for further exploration to refine both the predictive capability of the model and our understanding of loan applicants’ situations.

In our work, we have thoroughly documented the model's architecture as well as the preprocessing, fine-tuning, and training procedures. We have also used SHAP values to understand how the final model classifies and even how it makes decisions for individual instances. While these aspects are crucial in financial applications, a key limitation remains: the lack of transparency in how the BERT-based risk score is generated, which limits the understanding of the factors influencing the score and its potential biases. This challenge hinders the practical application of this approach in real-world settings, where it is critical to understand the score generation process and ensure that it does not introduce biases related to gender, ethnicity, or financial inclusion. Therefore, enhancing the transparency and explainability of the BERT score is essential, not only for regulatory compliance but also to build trust among borrowers and lenders \cite{ROFIEG2019}.

In this respect, several approaches exist to better understand how neural-based Natural Language Processing models such as BERT work \cite{Madsen2022}. Intrinsic methods, such as inspecting attention weights, do not offer sufficient transparency or meaningful explanations for model predictions \cite{JainWallace2019, serrano-smith-2019-attention}. Still, it is possible to use surrogate models, such as LIME \cite{LIME2016} or \textit{anchor} rules \cite{ANCHORS2018}, to try to interpret BERT-based predictions, as has been done in fake news detection \cite{BERTfakenewsXAI_2021}. Such an approach can help to better understand how these models work and facilitate their adoption in real-world settings.

A different approach that could also be explored is the use of LLM-based topic modeling techniques, such as BERTopic \cite{Grootendorst2022}. It could help to identify ``risky topics'' by making use of the deep semantic understanding of LLMs. These topics could then be used as additional input variables in the granting model, improving both predictive performance and transparency in the decision-making process. Additionally, the embeddings from BERT or other LLMs could be used to generate interpretable topics from large, complex vocabularies, as demonstrated in other studies \cite{Dieng_TopicModelingEmbedding}. Applying such topic modeling to loan descriptions could help identify loans with varying risk levels, further enhancing the transparency of credit risk assessments.

Further research could also explore the use of advanced LLMs. Encoder-only models like RoBERTa \cite{RoBERTa} may capture more intricate linguistic patterns while emerging generative AI approaches could redefine how risk scoring is performed. Although these models were not originally designed for such tasks, recent advancements have shown promising approximations. For instance, techniques like CARP (Clue And Reasoning Prompting) \cite{CARP2023} utilize in-context learning with few-shot examples to perform classification without fine-tuning. Applying such methods to risk scoring may open new possibilities for achieving robust results while minimizing computational overhead.

Future work should also address the economic implications of our findings by integrating cost- or profit-sensitive approaches \cite{Xia2017, Miller2024}. Investigating how the inclusion of textual descriptions impacts financial outcomes could provide valuable insights into the practical utility of these methods. By linking improved prediction performance to tangible economic benefits, we can further bridge the gap between academic innovation and real-world application.

\section*{Acknowledgements}
We thank Antonio Caparrini López and Miller Janny Ariza Garzón for their help at different stages of the project.

\bibliographystyle{plain}
\bibliography{references}

\end{document}